\documentclass{aa}
\usepackage{times}
\usepackage{txfonts}
\usepackage{graphics}
\usepackage{xspace}
\usepackage{epsfig}
\usepackage{natbib}
\usepackage{rotating}

\newcommand{\iras}{IRAS\,04108+2803\,B}

\begin{document}

\title{A statistical analysis of X-ray variability \\ in pre-main sequence objects of the Taurus Molecular Cloud}

\author{B. Stelzer\inst{1} \and E. Flaccomio\inst{1} \and K. Briggs\inst{2} \and G. Micela\inst{1} \and L. Scelsi\inst{3} \and M. Audard\inst{4} \and I. Pillitteri\inst{1,3} \and M. G\"udel\inst{2}}

\institute{INAF - Osservatorio Astronomico di Palermo,
  Piazza del Parlamento 1, I-90134 Palermo, Italy \and 
Paul Scherrer Institut, W\"urenlingen and Villigen, CH-5232 Villigen PSI, Switzerland \and
Dipartimento di Scienze Fisiche ed Astronomiche, Sezione di Astronomia, Universit\`a di Palermo,
Piazza del Parlamento 1, I-90134 Palermo, Italy \and
Columbia Astrophysics Laboratory, Mail Code 5247, 550 West 120th Street, New York, NY~10027, USA}

\offprints{B. Stelzer}
\mail{B. Stelzer, stelzer@astropa.unipa.it}
\titlerunning{X-ray variability on pre-MS stars in the TMC}

\date{Received $<$14-07-2006$>$ / Accepted $<$28-08-2006$>$} 

\abstract
{This work is part of a systematic X-ray survey of the Taurus star forming complex with XMM-Newton.}
{We study the time series of all X-ray sources associated with Taurus members, to statistically
characterize their X-ray variability, and compare the results to those for pre-main sequence stars
in the Orion Nebula Cluster and to expectations arising from a model where all the X-ray emission 
is the result of a large number of stochastically occurring flares.} 
{The analysis of the lightcurves is based on a maximum likelihood algorithm that segments the time series in 
intervals of constant signal without the need of binning. 
Flares are defined with criteria that take into account the amplitude
and the derivative of the segmented lightcurves.
Variability statistics
are evaluated for different classes of pre-main sequence stars (protostars, cTTS, wTTS, brown dwarfs),  
and for different spectral type ranges. Flare frequency and energy distribution are computed.}
{We find that roughly half of the detected X-ray sources show variability above our sensitivity limit,
and in $\sim 26$\,\% of the cases this variability is recognized as flares. Variability is more
frequently detected at hard than at soft energies. The variability statistics of cTTS and wTTS are 
undistinguishable, suggesting a common (coronal) origin for their X-ray emission. 
The frequency of large flares ($E > 10^{35}$\,erg) on Taurus members is $1$ event per star in $800$\,ksec. 
The typical duration of these flares -- probably biased by the finite observing time -- is 
about $10$\,ksec. 
We have for the first time applied a rigorous maximum likelihood method in the analysis of the 
number distribution of flare energies on pre-main sequence stars. 
In its differential form this distribution follows a power-law with index $\alpha = 2.4 \pm 0.5$, 
in the range typically observed on late-type stars and the Sun.
} 
{The signature of the X-ray variability in the pre-main sequence stars in Taurus and Orion 
provides twofold support for coronal heating by flares: 
(i) The correlation between the maximum variability amplitude and the minimum emission level 
indicates that both flare and quiescent emission are closely related to the coronal heating process. 
(ii) The power-law index $\alpha$ derived for the flare energy distribution is large enough to explain 
the heating of stellar coronae by nano-flares ($\alpha > 2$), albeit associated with a rather 
large uncertainty that leaves some doubt on this conclusion. 
}


\maketitle

\section{Introduction}\label{sect:intro}

X-ray variability is a characteristic property of magnetically 
active stars, including such diverse objects as the Sun, 
dMe flare stars, RS\,CVn variables, and pre-main-sequence (pre-MS) 
T Tauri stars \citep{Feigelson99.1, Favata03.1, Guedel04.1}. 
Variability on long time-scales may comprise modulation 
by rotating active regions (with time-scales on the order of days corresponding to the rotation
period of the star) or even activity cycles (with expected time-scales on the order of years);
see \citet{Marino03.1, Favata04.1} for examples. 
Most of the observed variations have short time-scales (on the order of hours), 
are accompanied by an increase in electron temperature, 
and are therefore attributed to flares resulting from magnetic reconnection events. 

The intense X-ray
radiation from flares is ascribed to the filling of magnetic loops with
heated gas that has been driven from the chromospheric layers along
the magnetic field lines into the corona 
\citep[the `chromospheric evaporation scenario', e.g.,][]{Antonucci84.1}. 
Various authors have therefore
speculated that the overall X-ray emission seen in magnetically active
stars is in fact the result of a large number of stochastically
occurring flares. In analogy to the `microflare heating' hypothesis in solar physics 
\citep[see e.g.][]{Hudson91.1}, these events are believed to be distributed in energy such that 
the large majority releases only small amounts of energy, making their individual 
identification in light curves impossible while we only measure the integrated emission
\citep[e.g.,][]{Audard00.1, Guedel02.1, Guedel03.1}. 

Extensive studies yielding statistical properties of
X-ray variability for different classes of active stars have long been impeded, 
mostly due to observational restrictions deriving from insufficient monitoring time.  
Further limitations, besides sensitivity, are imposed by the poor temporal coverage 
of X-ray instruments
in low-Earth orbits that result in frequent occultations of the target by the satellite 
revolution. 
{\em XMM-Newton} and {\em Chandra} have opened the field for systematic 
investigations on time-scales typical for magnetic activity.  
This is due to the eccentric orbits of both satellites, enabling 
uninterrupted observations of up to $\sim 2$\,d. 
Even then, variability studies are difficult, because in practice it is nearly impossible
to maximize both sample size and exposure time per target.
An outstanding study in this respect is the nearly 13\,d long observation
of the Orion Nebula Cluster (ONC) by {\em Chandra}, termed the {\em Chandra} Orion Ultradeep Project (COUP). 
The COUP yielded not only the longest uninterrupted X-ray time series for pre-MS stars so far, 
but owing to the dense population of the ONC it allowed also to examine in a homogeneous way the X-ray variability
of the largest sample of pre-MS stars: More than $1600$ X-ray sources were detected in a single {\em Chandra} 
field, mostly members of the ONC. 
Amongst others, the COUP resulted in a systematic study of flares on `young Suns' \citep{Wolk05.1}, 
rotational modulation \citep{Flaccomio05.1}, the geometry of flare loops \citep{Favata05.1}, and
X-ray variability of hot stars \citep{Stelzer05.1}. 

In contrast to the ONC, the Taurus star forming complex spans a large angle on the sky 
\citep[see Fig.~1 in ][]{Guedel06.1}.
This implies that observing a significant fraction of its stellar members with a telescope operating
in pointing mode requires many different exposures. Realistically, the whole population can be
sampled only by an all-sky monitoring. \citet{Neuhaeuser95.1} have used the data obtained
during the {\em ROSAT} All-Sky Survey (RASS) to study the X-ray emission of the pre-MS stars in Taurus. 
However, the RASS consisted in snapshots of a few seconds each separated by $\sim 90$\,min,
corresponding to the Earth revolution of the {\em ROSAT} satellite, and is inappropriate for
timing studies. A systematic study of the X-ray variability of the pre-MS stars in Taurus-Auriga 
based on more than $100$ {\it pointed} {\em ROSAT} PSPC observations was presented by \citet{Stelzer00.1}. 
That investigation yielded the first estimate of flare rates for these stars, which were found in  
outburst during about $1$\,\% of the observing time. Various uncertainties compromise this
measurement, e.g. 
the {\em ROSAT} data set 
was 
(i) inhomogeneous in sensitivity due to the vastly different exposure times of the 
individual observations, 
(ii) affected by Earth blocks due to the satellite's revolution, and 
(iii) limited to soft energies.

Here we present a variability study for the pre-MS stars in the Taurus star forming
region based on extensive {\em XMM-Newton} observations.  
We discuss data from the {\em XMM-Newton Extended Survey of the Taurus Molecular Cloud} (XEST).
This project was devised to 
comprise observations with roughly uniform and uninterrupted exposures of $\sim 30 - 40$\,ksec.
Thus, it represents a major improvement with respect to previous X-ray data of the Taurus Molecular
Cloud (TMC). 
A further advantage is the extension of the hard energy band beyond $2$\,keV. 
A detailed description of the XEST observations is found in \citet{Guedel06.1}. 
Here we present a systematic time series analysis of the pre-MS population detected in the XEST. 

In Sect.~\ref{sect:analysis} the methods used in the analysis of the time series 
are described, and the results are summarized. 
Sect.~\ref{sect:var_quant} deals with a quantitative investigation of the lightcurves. 
The relatively short duration of the XEST observations imply that the most likely cause of 
the variability is flaring.
We introduce our way of defining flares and present the results from the flare detection 
process in that section. 
Variability statistics for different classes of pre-MS stars in the TMC are given 
in Sect.~\ref{sect:results_var}, including a discussion of observational biases. 
The frequency and the energy distributions of the detected flares are
examined in Sect.~\ref{sect:flare_tau} and Sect.~\ref{sect:flare_energy}. 
Sect.~\ref{sect:ctts_wtts} presents a detailed comparison of the variability 
characteristics of cTTS and wTTS.
We summarize our results in Sect.~\ref{sect:conclusions}.

\section{Time series analysis}\label{sect:analysis}

The analysis presented in this paper includes the TMC members of all {\em XMM-Newton} 
fields from the XEST, i.e. the observations from our survey with {\em XMM-Newton} and the observations 
added from the {\em XMM-Newton} archive \citep[see observing log in Table~1 of ][]{Guedel06.1}, 
except field No.\,1 (T\,Tau) and No.\,25 (AA\,Tau), that are dedicated to separate projects 
(G\"udel et al., in prep.; Grosso et al., in prep.).
Some fields are spatially overlapping, such that several stars are detected in more than one 
exposure. 
Especially, in XEST fields No.23 and No.24 the satellite was pointing at the same sky position and the 
two exposures are adjacent in time (separated by only $\sim 1$\,ksec). 
Therefore, their data was merged for the time series analysis. 
The total number of X-ray sources in all studied XEST fields 
that are identified with known TMC members
after merging of XEST-23 and XEST-24 is $126$.

The time series analysis is based on the EPIC/pn photon event lists 
for the individual {\rm X-ray sources}. There are three exceptions where we
resort to the MOS data:
(i) for XEST field No.26 because there is no EPIC/pn data, 
(ii) for sources that are outside the field-of-view of EPIC/pn, or 
(iii) sources located on a chip gap resulting in a major loss 
of photons in EPIC/pn.  

The photon extraction regions for source and background photons have been 
defined taking care to avoid contributions from adjacent sources. 
The procedure is described in detail by \citet{Guedel06.1}.
For each X-ray source detected in both XEST-23 and XEST-24 the event lists
of the source and of the background of both observations were merged. 
For a given source the original photon extraction areas in XEST-23 and in XEST-24
may be slightly different. To avoid introducing artificial variability we selected 
for each star the smaller of the two extraction regions for both exposures. 

The statistical methods we used are independent of data binning, and
are described below. The analysis was carried out in different energy bands:
$0.3-7.8$\,keV (broad), $0.3-1.0$\,keV (soft), and $1.0-7.8$\,keV (hard).

\subsection{Maximum likelihood blocks}\label{subsect:mlb}

We searched for variability in each photon time series using a maximum likelihood
algorithm. The technique was described by \citet{Wolk05.1},
who have used the same method for an X-ray variability study of a subsample 
of the ONC members detected in the COUP. 
It is derived from the Bayesian blocks described by \citet{Scargle98.1}. 
In brief, the time series is split into periods
of constant signal (so-called `segments' or `blocks') under the assumption of Poisson
noise, by searching iteratively for change points in the intensity level. 
In contrast to conventional lightcurve analysis this method works directly on
the sequence of photon arrival times without the need for binning, such that ambiguities 
introduced by the choice of bin size and bin start are avoided.  
The maximum likelihood block (MLB) algorithm has two free parameters, 
the minimum number of counts per segment ($N_{\rm min}$) and the confidence level ($CL$)  
for the established change points. 
The significance thresholds for the change points were determined for any given
$N_{\rm min}$ through extensive simulations of constant lightcurves (Flaccomio et al., in prep.).
The results presented throughout this paper are for $CL = 99.9$\,\% and $N_{\rm min} = 20$.
Now we proceed to a justification of this choice and a detailed description of the data analysis steps. 

While the background of {\em Chandra} is very low and can be approximated by a constant, 
in {\em XMM-Newton} observations the background is non-negligible and variable. 
This complicates the analysis.  
To take account of the background and its variability we have extended the method with
respect to the use for the COUP. In short, the principle of this procedure is to create a 
background subtracted `source-only' or `net source' events list, by removing individual photons from 
the events list of the source position. Subsequently, the variability tests can be applied to
this net source events list. 

In practice, for the treatment of the background we proceed for each source
events list and the associated background events list in the following way:
First, the MLB algorithm is applied to the time series for the background,
i.e. the background events list (`B') is divided into ML blocks by identifying change
points in the background level. The matter is then how to subtract the background from the 
events detected at the source position (which at this point comprise both source and 
background photons; = `S+B$^\prime$'). 
We first scale the segmented background to the extraction area of the source. 
This gives the expected number of background events at the source position for each
segment in which the background is constant. 
The respective number of photons is then subtracted from the `S+B$^\prime$' events list  
uniformly within each time-interval of constant background. 
The result is a background subtracted source-only events list (`S'), ready for
further analysis. 

The MLB algorithm is now applied to `S' to yield the segmented source lightcurve. 
When computing the errors of the count rate in each segment, the
(subtracted) number of background photons within the respective time segment 
is considered.  
Note that, obviously, the segments of `S' are different and independent 
from the segments of `B'.  

\begin{figure*}[t]
\begin{center}
\parbox{18cm}{
\parbox{9cm}{
\resizebox{8.5cm}{!}{\includegraphics{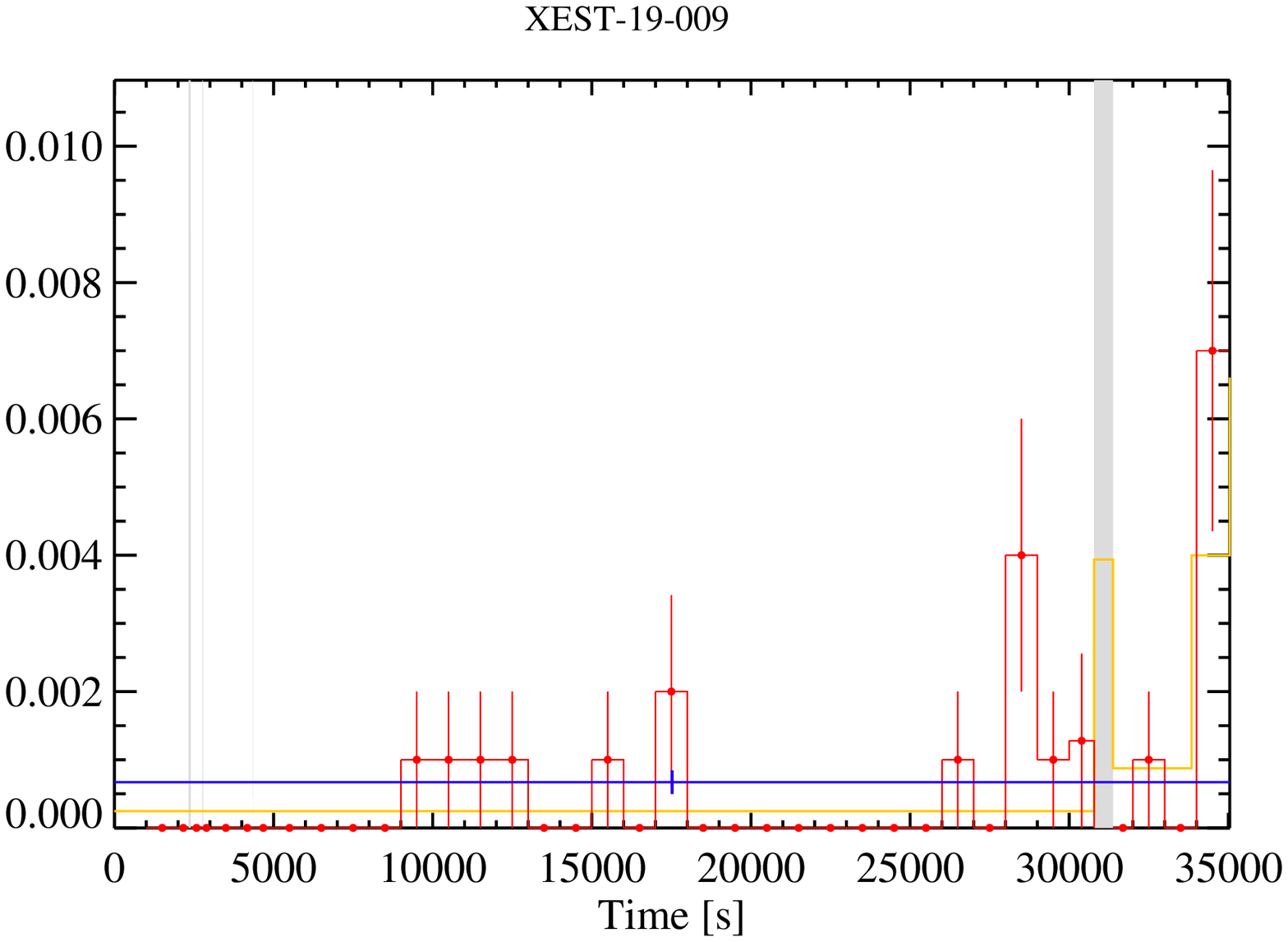}}
}
\parbox{9cm}{
\resizebox{8.5cm}{!}{\includegraphics{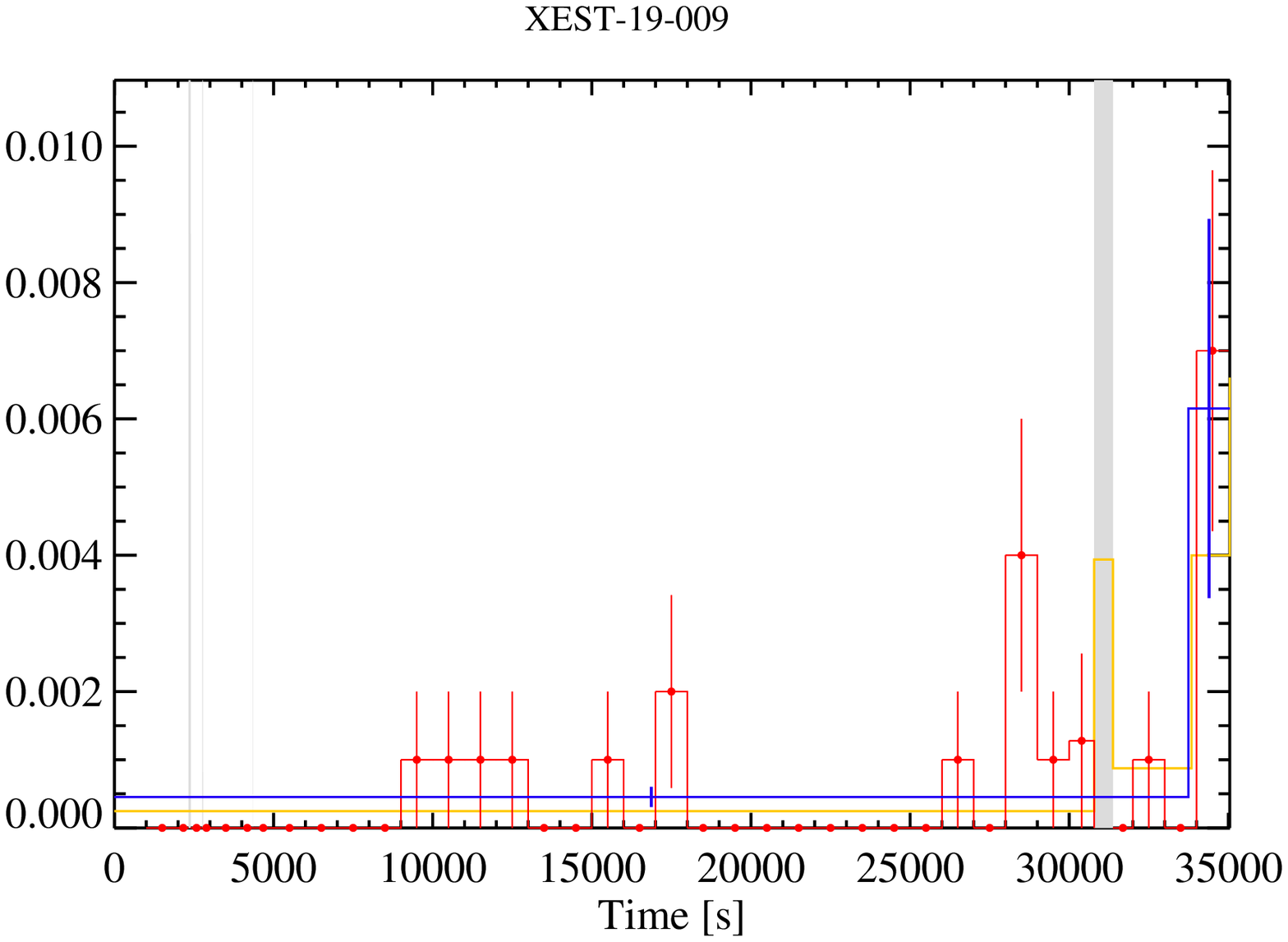}}
}
}
\caption{Broad band EPIC/MOS\,1 lightcurve of XEST-19-009. The background subtracted source signal is shown in two representations: binned into $1000$\,s intervals (red) and segmented with the MLB algorithm (blue). The left panel shows the segmentation obtained with $N_{\rm min}=20$ (only one block, i.e. constant lightcurve), the right panel shows the segmentation resulting from $N_{\rm min}=1$ (two blocks, i.e. variable lightcurve). The background segments scaled to the source area are shown in yellow. Time intervals rejected by our high-background filter are represented by grey-shaded areas. The division into two blocks with MLB\,1 probably results from imperfect background subtraction.}
\label{fig:lcs_mlb1_bad}
\end{center}
\end{figure*}
In general, a source can be considered variable at a certain confidence level 
if the background-subtracted blocked lightcurve consists of more than one segment. 
In the course of the analysis it was noticed that for some sources the detected 
variability coincides with times of very high and variable background. 
These variations could be a result of inaccurate background subtraction.
To check if the variations in these critical time intervals are spurious 
we simulated constant lightcurves by generating a random distribution of 
photon arrival times within a time interval corresponding to the length of the observation. 
The number of photons randomly generated was chosen such that the count rate of the
simulated data is equal to the average net count rate of the observed source. Then,
we superposed the observed area-scaled background on the simulated source events list, 
and carried out the background subtraction and MLB analysis. The procedure for
the treatment of the background is 
analogous to the one described above for the observed data, i.e.  
the background was subtracted blockwise from the simulated `${\rm S_{sim}+B}$' events list 
and a background subtracted simulated `${\rm S_{sim}}$' events list was computed. 

For each source $5000$ simulations were performed, with $CL=99.9$\,\% and $N_{\rm min}=20$. 
Then the blocked lightcurves were examined.
The simulated data sets are expected to yield no change point in the MLB analysis. 
However, in a substantial number of sources 
where the algorithm detected variability in the observed data,
variability was also detected in a high fraction ($\gg 10$\,\%) of the simulated lightcurves. 
These cases are found among faint sources that have high and strongly variable background,
and the statistical subtraction of background photons is problematic. 
 
To avoid such detections of spurious variability, 
for the ultimate analysis of the data we excluded all time intervals corresponding to blocks 
from `B' in which the number of events in the background after scaling to the source area 
is three or more times higher than the number of events in the source. 
In general, the reduction of the exposure time by
this `high-background filter' is low; less than $10$\,\% for $105$ of $126$ sources. The worst
cases occur in XEST-03 and XEST-24, where the background is so high, that -- depending on 
the source brightness -- up to $40$\,\% of the observing time was removed. 

Residual variability imposed by the background can be assessed with the help of the 
simulated data. 
Clearly, the effect of background induced variations is to decrease the significance of
any detected variability with respect to the chosen $CL$. 
In practice, we use the 
fraction of simulated event lists that appear non-variable with the MLB method 
to define a `corrected' confidence level ${CL}_{\rm corr}$ for the variability 
detected in the observed data.  
We delay the discussion of the results 
to Sect.~\ref{sect:results_var}, and proceed first with the description of the analysis steps.  

The analysis was performed in all three energy bands (soft, hard, and broad) 
with values for $N_{\rm min}$ of $1$ and $20$, henceforth 
referred to as MLB\,1 and MLB\,20 respectively.   
Higher values for $N_{\rm min}$ can be used e.g. to define time segments for time resolved 
spectral analysis \citep[see ][]{Franciosini06.1}. A low number of the allowed minimum 
counts per block favors the detection of variability in faint sources, but 
also the detection of spurious variations, that arise e.g. from residual background 
contamination.  
A lightcurve with an example for this latter effect is shown in Fig.~\ref{fig:lcs_mlb1_bad};
see figure caption for further explanations. 

A careful comparison showed that the variability statistics 
obtained with MLB\,1 and with MLB\,20 are widely consistent. 
%
%
For seven sources the segmentation of the broad band is different for MLB\,1 and MLB\,20. 
Among these, three sources are constant 
for $N_{\rm min} = 20$, but one change point is found with MLB\,1. 
Since these three sources are very faint, the detection of variability is hampered with
MLB\,20 (because the total number of net source counts is only 
slightly higher than the change point threshold $N_{\rm min}=20$). However, in practice, 
their variability detected with MLB\,1 is likely spurious because related to times of 
high background not rejected by
our high-background filter; see example in Fig.~\ref{fig:lcs_mlb1_bad}. 
For the other four sources at least one change point is detected with both MLB\,1 and MLB\,20,
and only the number or position of the change points is different between the two tests. 
To conclude, we find no case where the detection of variability is impeded with MLB\,20 due to the higher
threshold for the minimum number of counts per segment with respect to MLB\,1. 
Based on these results, we decided to present in this paper the results of the analysis 
with $N_{\rm min} = 20$.

\subsection{Kolmogorov-Smirnov test}\label{subsect:kstest}

As an additional investigation of the variability we performed a Kolmogorov-Smirnov (KS) test 
on the background
subtracted events lists. The output of the KS test is a probability $P_{\rm KS}$ that
the variability detected in the observed data is physical. 
Since the `S' events file has been manipulated by the subtraction
of the background, it is not immediately obvious that the KS-test is applicable. 
To examine the validity of the KS analysis we made use of the simulations of constant source 
event lists described above. Analogously to the MLB analysis, the simulated lightcurves
should be non-variable against the KS-test, except for spurious variability detections
due to random fluctuations in the photon arrival times and additional noise introduced by the 
background subtraction. To quantify the effect of such spurious variations, 
we computed for each variable source the fraction $f_{\rm P}$ of simulated lightcurves that 
show a probability $P_{\rm KS}$ of being variable higher than the value derived for the observed 
photon time series. The `effective' probability for variability in the data is 
$P_{\rm KS,eff} = 1-f_{\rm P}$.

\begin{sidewaystable*}\begin{center}\small
\caption{X-ray variability of TMC members during the XEST. Cols.~$1-4$ give the XEST number, stellar identification, young star class, and spectral type \protect\citep[see ][for details on these classifications]{Guedel06.1}. The summed broad band GTI after removal of time intervals with high background is given in col.~$5$. The analysis is based on EPIC/pn data; exceptions are marked with labels $M1$ for MOS\,1 and $M2$ for MOS\,2. The remaining columns represent the number of source photons (`Cts'), the number of segments (`$N_{\rm b}$'), the confidence level of the MLB test (`$CL_{\rm corr}$'), and the probability for variable source according to the KS-test (`$P_{\rm KS,eff}$'). Results are given for the broad ($0.3-7.8$\,keV), soft ($0.3-1.0$\,keV) and hard ($1.0-7.8$\,keV) band.}
\label{tab:var}
\begin{tabular}{llccrp{0.12cm}rrrrrrrrrrrr}\hline
     &             &      &     &          & & \multicolumn{4}{c}{-------- Broad band --------}                   & \multicolumn{4}{c}{-------- Soft band --------}                    & \multicolumn{4}{c}{-------- Hard band --------}                    \\
XEST No. & Identification & Type & SpT & Expo [s] & & Cts & $N_{\rm b}$ & $CL_{\rm corr}$ & $P_{\rm KS,eff}$ & Cts & $N_{\rm b}$ & $CL_{\rm corr}$ & $P_{\rm KS,eff}$ & Cts & $N_{\rm b}$ & $CL_{\rm corr}$ & $P_{\rm KS,eff}$ \\
\hline\hline
27-115 &  HBC 352             & $       3$ & G0        & $  39056$ & 
\hspace*{0.1cm} & $1601$ & $2$ & $0.93$ & $1.00$ & $842$ & $2$ & $0.96$ & $
1.00$ & $772$ & $3$ & $0.93$ & $1.00$ \\
06-005 &  HBC 358 AB          & $       3$ & M2        & $  29919$ & 
\hspace*{0.1cm} & $451$ & $1$ & $    $ & $0.80$ & $328$ & $1$ & $    $ & $0.93
$ & $122$ & $1$ & $    $ & $0.34$ \\
06-007 &  HBC 359             & $       3$ & M2        & $  29918$ & 
\hspace*{0.1cm} & $1091$ & $1$ & $    $ & $0.30$ & $791$ & $1$ & $    $ & $
0.83$ & $304$ & $1$ & $    $ & $0.64$ \\
06-059 &  L1489 IRS           & $       1$ & K4        & $  29804$ & 
\hspace*{0.1cm} & $1537$ & $1$ & $    $ & $0.98$ & $8$ & $1$ & $    $ & $0.63
$ & $1537$ & $1$ & $    $ & $0.98$ \\
20-001 &  LkCa 1              & $       3$ & M4        & $  29872$ & 
\hspace*{0.1cm} & $318$ & $1$ & $    $ & $0.93$ & $260$ & $1$ & $    $ & $0.90
$ & $58$ & $1$ & $    $ & $0.83$ \\
20-005 &  Anon 1              & $       3$ & M0        & $  29872$ & 
\hspace*{0.1cm} & $4523$ & $3$ & $0.98$ & $1.00$ & $1973$ & $1$ & $    $ & $
0.97$ & $2550$ & $3$ & $0.98$ & $1.00$ \\
20-022 &  IRAS 04108+2803 B   & $       1$ &           & $  14342$ & 
\hspace*{0.1cm} & $368$ & $3$ & $0.97$ & $1.00$ & $-$ & $-$ & $   -$ & $   -
$ & $371$ & $3$ & $0.97$ & $1.00$ \\
20-042 &  V773 Tau ABC        & $       3$ & K2/M0     & $  29867$ & 
\hspace*{0.1cm} & $29820$ & $1$ & $    $ & $0.98$ & $15546$ & $1$ & $    $ & $
0.49$ & $14274$ & $2$ & $0.99$ & $1.00$ \\
20-043 &  FM Tau              & $       2$ & M0        & $  29876$ & 
\hspace*{0.1cm} & $1142$ & $2$ & $0.96$ & $1.00$ & $553$ & $2$ & $0.95$ & $
0.99$ & $589$ & $2$ & $0.95$ & $1.00$ \\
20-046 &  CW Tau              & $       2$ & K3        & $  29877$ & 
\hspace*{0.1cm} & $32$ & $1$ & $    $ & $0.75$ & $-$ & $-$ & $   -$ & $   -
$ & $28$ & $1$ & $    $ & $0.90$ \\
20-047 &  CIDA 1              & $       2$ & M5.5      & $  29877$ & 
\hspace*{0.1cm} & $30$ & $1$ & $    $ & $0.41$ & $20$ & $1$ & $    $ & $0.44
$ & $10$ & $1$ & $    $ & $0.92$ \\
20-056 &  MHO 2/1             & $       2$ & M2.5/2.5  & $  29877$ & 
\hspace*{0.1cm} & $1298$ & $4$ & $0.98$ & $1.00$ & $117$ & $1$ & $    $ & $
0.99$ & $1182$ & $2$ & $0.97$ & $1.00$ \\
20-058 &  MHO 3               & $       2$ & K7        & $  29877$ & 
\hspace*{0.1cm} & $332$ & $1$ & $    $ & $0.97$ & $84$ & $1$ & $    $ & $0.12
$ & $249$ & $1$ & $    $ & $0.97$ \\
20-069 &  FO Tau AB           & $       2$ & M2        & $  29873$ & 
\hspace*{0.1cm} & $105$ & $1$ & $    $ & $0.97$ & $44$ & $1$ & $    $ & $0.78
$ & $61$ & $1$ & $    $ & $0.93$ \\
20-073 &  CIDA 2              & $       3$ & M5.5      & $  29874$ & 
\hspace*{0.1cm} & $319$ & $1$ & $    $ & $0.50$ & $241$ & $1$ & $    $ & $0.10
$ & $78$ & $1$ & $    $ & $0.48$ \\
23-002/24-002 &  CY Tau              & $       2$ & M1.5      & $ 113136$ & 
\hspace*{0.05cm}$^{M1}$ & $371$ & $4$ & $0.94$ & $1.00$ & $196$ & $1$ & $    
$ & $0.99$ & $171$ & $3$ & $0.92$ & $1.00$ \\
23-004/24-004 &  LkCa 5              & $       3$ & M2        & $ 106682$ & 
\hspace*{0.1cm} & $4551$ & $5$ & $0.90$ & $1.00$ & $3101$ & $3$ & $0.98$ & $
1.00$ & $1459$ & $5$ & $0.75$ & $1.00$ \\
23-008/24-008 &  CIDA 3              & $       3$ & M4        & $  91664$ & 
\hspace*{0.1cm} & $349$ & $2$ & $0.77$ & $1.00$ & $19$ & $1$ & $    $ & $0.99
$ & $347$ & $2$ & $0.84$ & $1.00$ \\
23-015/24-015 &  V410 X3             & $       3$ & M6        & $  97016$ & 
\hspace*{0.1cm} & $597$ & $1$ & $    $ & $0.09$ & $439$ & $1$ & $    $ & $0.36
$ & $166$ & $1$ & $    $ & $0.98$ \\
23-018 &  V410 A13            & $       2$ & M5.75     & $  53298$ & 
\hspace*{0.1cm} & $23$ & $1$ & $    $ & $0.93$ & $7$ & $1$ & $    $ & $0.90
$ & $14$ & $1$ & $    $ & $0.83$ \\
23-029/24-027 &  V410 A25            & $       9$ & M1        & $ 108957$ & 
\hspace*{0.1cm} & $3211$ & $2$ & $0.87$ & $1.00$ & $345$ & $1$ & $    $ & $
0.19$ & $2871$ & $2$ & $0.92$ & $1.00$ \\
23-032/24-028 &  V410 Tau ABC        & $       3$ & K4        & $ 107686$ & 
\hspace*{0.1cm} & $116409$ & $16$ & $0.98$ & $1.00$ & $74323$ & $12$ & $0.98
$ & $1.00$ & $42079$ & $14$ & $0.96$ & $1.00$ \\
23-033/24-029 &  DD Tau AB           & $       2$ & M3        & $  99316$ & 
\hspace*{0.1cm} & $594$ & $4$ & $0.89$ & $1.00$ & $120$ & $1$ & $    $ & $0.68
$ & $481$ & $5$ & $0.86$ & $1.00$ \\
23-035/24-030 &  CZ Tau AB           & $       3$ & M3        & $ 106468$ & 
\hspace*{0.1cm} & $1372$ & $2$ & $0.92$ & $1.00$ & $987$ & $2$ & $0.97$ & $
1.00$ & $385$ & $3$ & $0.84$ & $1.00$ \\
23-036/24-031 &  IRAS 04154+2823     & $       2$ & M2.5      & $  85864$ & 
\hspace*{0.1cm} & $191$ & $2$ & $0.67$ & $1.00$ & $17$ & $1$ & $    $ & $0.75
$ & $178$ & $2$ & $0.74$ & $1.00$ \\
23-037/24-032 &  V410 X2             & $       9$ & M0        & $ 103309$ & 
\hspace*{0.1cm} & $3625$ & $4$ & $0.73$ & $1.00$ & $475$ & $2$ & $0.78$ & $
0.77$ & $3133$ & $3$ & $0.41$ & $1.00$ \\
23-045/24-038 &  V410 X4             & $       9$ & M4        & $  94996$ & 
\hspace*{0.1cm} & $810$ & $3$ & $0.60$ & $1.00$ & $89$ & $1$ & $    $ & $0.71
$ & $749$ & $3$ & $0.73$ & $1.00$ \\
23-047/24-040 &  V892 Tau            & $       5$ & B9        & $ 109954$ & 
\hspace*{0.1cm} & $28548$ & $11$ & $0.97$ & $1.00$ & $2354$ & $2$ & $0.96$ & $
1.00$ & $26196$ & $13$ & $0.97$ & $1.00$ \\
23-048 &  LR 1                & $       9$ & K4.5      & $  43482$ & 
\hspace*{0.1cm} & $68$ & $2$ & $0.92$ & $1.00$ & $-$ & $-$ & $   -$ & $   -
$ & $83$ & $1$ & $    $ & $0.99$ \\
23-050/24-042 &  V410 X7             & $       3$ & M0.75     & $ 110155$ & 
\hspace*{0.1cm} & $3259$ & $6$ & $0.91$ & $1.00$ & $413$ & $3$ & $0.91$ & $
1.00$ & $2846$ & $6$ & $0.93$ & $1.00$ \\
23-056/24-047 &  Hubble 4            & $       3$ & K7        & $ 109845$ & 
\hspace*{0.1cm} & $30574$ & $4$ & $0.96$ & $1.00$ & $14776$ & $1$ & $    $ & $
0.66$ & $15791$ & $7$ & $0.95$ & $1.00$ \\
23-061/24-054 &  V410 X6             & $       3$ & M5.5      & $  97683$ & 
\hspace*{0.1cm} & $436$ & $2$ & $0.84$ & $1.00$ & $222$ & $2$ & $0.88$ & $1.00
$ & $207$ & $2$ & $0.82$ & $1.00$ \\
23-063/24-055 &  V410 X5             & $       3$ & M5.5      & $ 104095$ & 
\hspace*{0.1cm} & $2058$ & $3$ & $0.88$ & $1.00$ & $631$ & $2$ & $0.60$ & $
1.00$ & $1439$ & $3$ & $0.76$ & $1.00$ \\
23-067/24-058 &  FQ Tau AB           & $       2$ & M3/M3.5   & $  95824$ & 
\hspace*{0.05cm}$^{M1}$ & $75$ & $1$ & $    $ & $0.79$ & $39$ & $1$ & $    
$ & $0.90$ & $37$ & $1$ & $    $ & $0.35$ \\
28-100 &  BP Tau              & $       2$ & K7        & $ 129661$ & 
\hspace*{0.1cm} & $28835$ & $12$ & $0.91$ & $1.00$ & $15671$ & $8$ & $0.97
$ & $1.00$ & $13126$ & $12$ & $0.93$ & $1.00$ \\
23-074/24-061 &  V819 Tau AB         & $       3$ & K7        & $ 106945$ & 
\hspace*{0.1cm} & $9857$ & $4$ & $0.85$ & $1.00$ & $6132$ & $3$ & $0.95$ & $
1.00$ & $3699$ & $6$ & $0.70$ & $1.00$ \\
11-023 &  2M J04213459        & $       3$ & M5.5      & $  31441$ & 
\hspace*{0.1cm} & $77$ & $1$ & $    $ & $0.94$ & $53$ & $1$ & $    $ & $0.85
$ & $26$ & $1$ & $    $ & $0.04$ \\
11-037 &  CFHT-Tau 10         & $       3$ & M5.75     & $  28006$ & 
\hspace*{0.1cm} & $10$ & $1$ & $    $ & $0.59$ & $10$ & $1$ & $    $ & $0.46
$ & $-$ & $-$ & $   -$ & $   -$ \\
21-038 &  RY Tau              & $       2$ & K1        & $  45030$ & 
\hspace*{0.1cm} & $6864$ & $3$ & $0.91$ & $1.00$ & $991$ & $2$ & $0.87$ & $
1.00$ & $5877$ & $2$ & $0.92$ & $1.00$ \\
21-039 &  HD 283572           & $       3$ & G5        & $  44346$ & 
\hspace*{0.1cm} & $94102$ & $4$ & $0.98$ & $1.00$ & $55032$ & $2$ & $0.98$ & $
1.00$ & $39073$ & $4$ & $0.97$ & $1.00$ \\
\hline
\end{tabular}\end{center}
\end{sidewaystable*}

\addtocounter{table}{-1}

\begin{sidewaystable*}\begin{center}\small
\caption{{\em continued}}
\begin{tabular}{llccrp{0.12cm}rrrrrrrrrrrr}\hline
     &             &      &     &          & & \multicolumn{4}{c}{-------- Broad band --------}                   & \multicolumn{4}{c}{-------- Soft band --------}                    & \multicolumn{4}{c}{-------- Hard band --------}                    \\
XEST No. & Identification & Type & SpT & Expo [s] & & Cts & $N_{\rm b}$ & $CL_{\rm corr}$ & $P_{\rm KS,eff}$ & Cts & $N_{\rm b}$ & $CL_{\rm corr}$ & $P_{\rm KS,eff}$ & Cts & $N_{\rm b}$ & $CL_{\rm corr}$ & $P_{\rm KS,eff}$ \\
\hline\hline
11-054 &  Haro 6-5 B          & $       1$ & K5        & $  35463$ & 
\hspace*{0.1cm} & $64$ & $1$ & $    $ & $0.69$ & $6$ & $1$ & $    $ & $0.96
$ & $59$ & $1$ & $    $ & $0.77$ \\
11-057 &  FS Tau AC           & $       2$ & M0/M3.5   & $  37353$ & 
\hspace*{0.1cm} & $2845$ & $5$ & $0.97$ & $1.00$ & $120$ & $1$ & $    $ & $
0.28$ & $2729$ & $5$ & $0.97$ & $1.00$ \\
21-044 &  LkCa 21             & $       3$ & M3        & $  45474$ & 
\hspace*{0.1cm} & $3539$ & $2$ & $0.88$ & $1.00$ & $2467$ & $1$ & $    $ & $
0.62$ & $892$ & $2$ & $0.80$ & $1.00$ \\
11-079 &  CFHT-Tau 21         & $       2$ & M1.25     & $  34673$ & 
\hspace*{0.1cm} & $164$ & $2$ & $0.92$ & $1.00$ & $9$ & $1$ & $    $ & $0.88
$ & $141$ & $2$ & $0.95$ & $0.99$ \\
02-013 &  FV Tau AB           & $       2$ & K5/K6     & $  38702$ & 
\hspace*{0.1cm} & $310$ & $1$ & $    $ & $0.99$ & $7$ & $1$ & $    $ & $0.90
$ & $300$ & $1$ & $    $ & $0.99$ \\
02-016 &  KPNO-Tau 13         & $       3$ & M5        & $  38190$ & 
\hspace*{0.1cm} & $279$ & $1$ & $    $ & $0.78$ & $122$ & $1$ & $    $ & $0.68
$ & $153$ & $1$ & $    $ & $0.99$ \\
02-022 &  DG Tau A            & $       2$ & K6        & $  39027$ & 
\hspace*{0.1cm} & $681$ & $3$ & $0.94$ & $1.00$ & $267$ & $1$ & $    $ & $0.78
$ & $384$ & $3$ & $0.95$ & $1.00$ \\
15-020 &  JH 507              & $       3$ & M4        & $  27652$ & 
\hspace*{0.1cm} & $896$ & $2$ & $0.91$ & $0.99$ & $632$ & $2$ & $0.96$ & $1.00
$ & $260$ & $1$ & $    $ & $0.98$ \\
13-004 &  GV Tau AB           & $       1$ & K3-7      & $  26617$ & 
\hspace*{0.1cm} & $296$ & $3$ & $0.88$ & $1.00$ & $50$ & $1$ & $    $ & $0.42
$ & $248$ & $3$ & $0.90$ & $1.00$ \\
15-040 &  DH Tau AB           & $       2$ & M1        & $  30891$ & 
\hspace*{0.1cm} & $16216$ & $5$ & $0.98$ & $1.00$ & $7840$ & $4$ & $0.98$ & $
1.00$ & $8375$ & $5$ & $0.97$ & $1.00$ \\
15-042 &  DI Tau AB           & $       3$ & M0        & $  30891$ & 
\hspace*{0.1cm} & $2049$ & $1$ & $    $ & $0.98$ & $1340$ & $1$ & $    $ & $
0.60$ & $693$ & $2$ & $0.82$ & $1.00$ \\
15-044 &  KPNO-Tau 5          & $       4$ & M7.5      & $  18419$ & 
\hspace*{0.1cm} & $40$ & $1$ & $    $ & $0.90$ & $45$ & $1$ & $    $ & $0.70
$ & $22$ & $1$ & $    $ & $0.93$ \\
14-006 &  IQ Tau A            & $       2$ & M0.5      & $  30377$ & 
\hspace*{0.1cm} & $1079$ & $3$ & $0.94$ & $1.00$ & $179$ & $2$ & $0.95$ & $
1.00$ & $903$ & $3$ & $0.95$ & $1.00$ \\
13-035 &  FX Tau AB           & $       2$ & M1        & $  30547$ & 
\hspace*{0.05cm}$^{M2}$ & $162$ & $1$ & $    $ & $0.94$ & $81$ & $1$ & $    
$ & $0.90$ & $81$ & $1$ & $    $ & $0.89$ \\
14-057 &  DK Tau AB           & $       2$ & K7        & $  31879$ & 
\hspace*{0.1cm} & $2358$ & $1$ & $    $ & $0.28$ & $1021$ & $1$ & $    $ & $
0.67$ & $1329$ & $1$ & $    $ & $0.05$ \\
22-013 &  MHO 9               & $       3$ & M4.25     & $  52864$ & 
\hspace*{0.1cm} & $355$ & $1$ & $    $ & $0.98$ & $287$ & $1$ & $    $ & $0.95
$ & $77$ & $1$ & $    $ & $0.80$ \\
22-021 &  MHO 4               & $       4$ & M7.1      & $  53756$ & 
\hspace*{0.1cm} & $313$ & $1$ & $    $ & $0.53$ & $209$ & $1$ & $    $ & $0.86
$ & $105$ & $1$ & $    $ & $0.54$ \\
22-040 &  L1551 IRS5          & $       1$ &           & $  51467$ & 
\hspace*{0.1cm} & $84$ & $1$ & $    $ & $0.77$ & $31$ & $1$ & $    $ & $0.05
$ & $57$ & $1$ & $    $ & $0.85$ \\
22-042 &  LkHa 358            & $       2$ & M5.5      & $  52730$ & 
\hspace*{0.1cm} & $70$ & $2$ & $0.63$ & $1.00$ & $-$ & $-$ & $   -$ & $   -
$ & $81$ & $2$ & $0.56$ & $0.84$ \\
22-043 &  HL Tau              & $       1$ & K5        & $  54434$ & 
\hspace*{0.1cm} & $1416$ & $2$ & $0.97$ & $0.99$ & $5$ & $1$ & $    $ & $0.88
$ & $1409$ & $2$ & $0.98$ & $0.99$ \\
22-047 &  XZ Tau AB           & $       2$ & M2/M3.5   & $  54434$ & 
\hspace*{0.1cm} & $8181$ & $5$ & $0.97$ & $1.00$ & $2894$ & $4$ & $0.98$ & $
1.00$ & $5286$ & $6$ & $0.97$ & $1.00$ \\
22-056 &  L1551 NE            & $       1$ &           & $  52096$ & 
\hspace*{0.1cm} & $16$ & $1$ & $    $ & $0.72$ & $-$ & $-$ & $   -$ & $   -
$ & $20$ & $1$ & $    $ & $0.72$ \\
03-005 &  HK Tau AB           & $       2$ & M0.5/M2   & $  19581$ & 
\hspace*{0.1cm} & $45$ & $2$ & $0.92$ & $0.98$ & $14$ & $1$ & $    $ & $0.99
$ & $27$ & $1$ & $    $ & $0.71$ \\
22-070 &  V710 Tau BA         & $       2$ & M0.5/M2   & $  54346$ & 
\hspace*{0.1cm} & $3191$ & $1$ & $    $ & $0.82$ & $1611$ & $1$ & $    $ & $
0.79$ & $1569$ & $1$ & $    $ & $0.19$ \\
19-009 &  JH 665              & $       3$ & M5.5      & $  34356$ & 
\hspace*{0.05cm}$^{M1}$ & $23$ & $1$ & $    $ & $0.99$ & $7$ & $1$ & $    
$ & $0.60$ & $19$ & $1$ & $    $ & $1.00$ \\
22-089 &  L1551 51            & $       3$ & K7        & $  54443$ & 
\hspace*{0.1cm} & $4739$ & $4$ & $0.97$ & $1.00$ & $3127$ & $3$ & $0.98$ & $
1.00$ & $1603$ & $3$ & $0.95$ & $1.00$ \\
22-097 &  V827 Tau            & $       3$ & K7        & $  56844$ & 
\hspace*{0.05cm}$^{M1}$ & $5255$ & $2$ & $0.98$ & $1.00$ & $2593$ & $2$ & $
0.98$ & $1.00$ & $2661$ & $2$ & $0.98$ & $1.00$ \\
03-016 &  Haro 6-13           & $       2$ & M0        & $  23372$ & 
\hspace*{0.1cm} & $203$ & $2$ & $0.86$ & $0.99$ & $27$ & $1$ & $    $ & $0.28
$ & $188$ & $2$ & $0.88$ & $1.00$ \\
22-100 &  V826 Tau            & $       3$ & K7        & $  54443$ & 
\hspace*{0.1cm} & $18357$ & $3$ & $0.98$ & $1.00$ & $11587$ & $2$ & $0.98$ & $
1.00$ & $6767$ & $2$ & $0.97$ & $0.99$ \\
22-101 &  MHO 5               & $       2$ & M6        & $  52251$ & 
\hspace*{0.1cm} & $533$ & $1$ & $    $ & $0.80$ & $418$ & $1$ & $    $ & $0.46
$ & $124$ & $1$ & $    $ & $0.21$ \\
03-017 &  CFHT-Tau 7          & $       3$ & M5.75     & $  32583$ & 
\hspace*{0.1cm} & $104$ & $1$ & $    $ & $0.92$ & $81$ & $1$ & $    $ & $0.61
$ & $10$ & $1$ & $    $ & $0.82$ \\
03-019 &  V928 Tau AB         & $       3$ & M0.5      & $  32024$ & 
\hspace*{0.1cm} & $1545$ & $1$ & $    $ & $0.97$ & $826$ & $1$ & $    $ & $
0.98$ & $721$ & $1$ & $    $ & $0.80$ \\
03-022 &  FY Tau              & $       2$ & K5        & $  33093$ & 
\hspace*{0.1cm} & $1271$ & $4$ & $0.96$ & $0.99$ & $362$ & $1$ & $    $ & $
0.89$ & $910$ & $3$ & $0.96$ & $1.00$ \\
03-023 &  FZ Tau              & $       2$ & M0        & $  34957$ & 
\hspace*{0.05cm}$^{M1}$ & $117$ & $1$ & $    $ & $0.99$ & $54$ & $1$ & $    
$ & $0.95$ & $63$ & $1$ & $    $ & $0.97$ \\
17-002 &  IRAS 04295+2251     & $       1$ &           & $  27556$ & 
\hspace*{0.1cm} & $149$ & $2$ & $0.95$ & $1.00$ & $7$ & $1$ & $    $ & $0.97
$ & $140$ & $2$ & $0.95$ & $1.00$ \\
19-049 &  UZ Tau E+W(AB)      & $       2$ & M1/2/3    & $  32225$ & 
\hspace*{0.1cm} & $1292$ & $2$ & $0.92$ & $0.91$ & $599$ & $1$ & $    $ & $
0.44$ & $704$ & $2$ & $0.91$ & $0.77$ \\
17-009 &  JH 112              & $       2$ & K6        & $  28498$ & 
\hspace*{0.1cm} & $453$ & $1$ & $    $ & $0.99$ & $143$ & $1$ & $    $ & $0.62
$ & $307$ & $1$ & $    $ & $0.99$ \\
03-031 &  CFHT-Tau 5          & $       4$ & M7.5      & $  32755$ & 
\hspace*{0.05cm}$^{M1}$ & $37$ & $1$ & $    $ & $0.83$ & $6$ & $1$ & $    
$ & $0.54$ & $28$ & $1$ & $    $ & $0.96$ \\
04-003 &  CFHT-Tau 5          & $       4$ & M7.5      & $  29570$ & 
\hspace*{0.1cm} & $124$ & $1$ & $    $ & $0.72$ & $25$ & $1$ & $    $ & $0.30
$ & $99$ & $1$ & $    $ & $0.92$ \\
03-035 &  MHO 8               & $       3$ & M6        & $  21097$ & 
\hspace*{0.1cm} & $78$ & $1$ & $    $ & $0.99$ & $73$ & $1$ & $    $ & $0.81
$ & $41$ & $2$ & $0.93$ & $1.00$ \\
\hline
\end{tabular}\end{center}
\end{sidewaystable*}

\addtocounter{table}{-1}

\begin{sidewaystable*}\begin{center}\small
\caption{{\em continued}}
\begin{tabular}{llccrp{0.12cm}rrrrrrrrrrrr}\hline
     &             &      &     &          & & \multicolumn{4}{c}{-------- Broad band --------}                   & \multicolumn{4}{c}{-------- Soft band --------}                    & \multicolumn{4}{c}{-------- Hard band --------}                    \\
XEST No. & Identification & Type & SpT & Expo [s] & & Cts & $N_{\rm b}$ & $CL_{\rm corr}$ & $P_{\rm KS,eff}$ & Cts & $N_{\rm b}$ & $CL_{\rm corr}$ & $P_{\rm KS,eff}$ & Cts & $N_{\rm b}$ & $CL_{\rm corr}$ & $P_{\rm KS,eff}$ \\
\hline\hline
04-009 &  MHO 8               & $       3$ & M6        & $  29561$ & 
\hspace*{0.1cm} & $73$ & $1$ & $    $ & $0.63$ & $54$ & $1$ & $    $ & $0.78
$ & $20$ & $1$ & $    $ & $0.18$ \\
04-010 &  GH Tau AB           & $       2$ & M1.5/M2   & $  33077$ & 
\hspace*{0.05cm}$^{M1}$ & $42$ & $1$ & $    $ & $0.96$ & $10$ & $1$ & $    
$ & $0.78$ & $32$ & $1$ & $    $ & $0.93$ \\
04-012 &  V807 Tau (SNab)     & $       2$ & K7/M3     & $  30317$ & 
\hspace*{0.1cm} & $1733$ & $3$ & $0.98$ & $1.00$ & $1207$ & $1$ & $    $ & $
0.94$ & $525$ & $3$ & $0.98$ & $1.00$ \\
18-004 &  KPNO-Tau 14         & $       3$ & M6        & $  27895$ & 
\hspace*{0.1cm} & $481$ & $3$ & $0.94$ & $1.00$ & $62$ & $1$ & $    $ & $0.85
$ & $418$ & $3$ & $0.96$ & $1.00$ \\
04-016 &  V830 Tau            & $       3$ & K7        & $  30411$ & 
\hspace*{0.1cm} & $9455$ & $5$ & $0.89$ & $1.00$ & $5618$ & $4$ & $0.98$ & $
1.00$ & $3839$ & $5$ & $0.83$ & $1.00$ \\
17-027 &  IRAS 04303+2240     & $       2$ & M0.5      & $  29039$ & 
\hspace*{0.1cm} & $2257$ & $3$ & $0.96$ & $0.95$ & $114$ & $2$ & $0.92$ & $
1.00$ & $2148$ & $1$ & $    $ & $0.88$ \\
04-034 &  GI Tau              & $       2$ & K7        & $  30814$ & 
\hspace*{0.1cm} & $748$ & $1$ & $    $ & $0.86$ & $269$ & $1$ & $    $ & $0.91
$ & $480$ & $1$ & $    $ & $0.62$ \\
04-035 &  GK Tau AB           & $       2$ & K7        & $  30814$ & 
\hspace*{0.1cm} & $1334$ & $2$ & $0.96$ & $1.00$ & $481$ & $1$ & $    $ & $
0.79$ & $853$ & $2$ & $0.96$ & $1.00$ \\
18-019 &  IS Tau AB           & $       2$ & K7/M4.5   & $  27970$ & 
\hspace*{0.1cm} & $1000$ & $2$ & $0.97$ & $1.00$ & $318$ & $1$ & $    $ & $
0.92$ & $683$ & $2$ & $0.97$ & $1.00$ \\
17-058 &  CI Tau              & $       2$ & K7        & $  27575$ & 
\hspace*{0.1cm} & $243$ & $2$ & $0.94$ & $1.00$ & $40$ & $2$ & $0.98$ & $0.98
$ & $204$ & $2$ & $0.95$ & $1.00$ \\
18-030 &  IT Tau AB           & $       2$ & K2        & $  28666$ & 
\hspace*{0.1cm} & $9304$ & $2$ & $0.98$ & $1.00$ & $1102$ & $1$ & $    $ & $
0.56$ & $8194$ & $2$ & $0.98$ & $1.00$ \\
17-066 &  JH 108              & $       3$ & M1        & $  28847$ & 
\hspace*{0.1cm} & $2539$ & $4$ & $0.97$ & $1.00$ & $993$ & $3$ & $0.97$ & $
1.00$ & $1546$ & $5$ & $0.95$ & $1.00$ \\
17-068 &  CFHT-BD Tau 1       & $       4$ & M7.1      & $  26906$ & 
\hspace*{0.1cm} & $86$ & $1$ & $    $ & $0.99$ & $13$ & $1$ & $    $ & $0.81
$ & $70$ & $1$ & $    $ & $0.99$ \\
09-010 &  HO Tau AB           & $       2$ & M0.5      & $  21904$ & 
\hspace*{0.1cm} & $61$ & $1$ & $    $ & $0.31$ & $52$ & $1$ & $    $ & $0.37
$ & $19$ & $1$ & $    $ & $0.82$ \\
08-019 &  FF Tau AB           & $       3$ & K7        & $  34927$ & 
\hspace*{0.1cm} & $1345$ & $3$ & $0.88$ & $1.00$ & $657$ & $2$ & $0.94$ & $
1.00$ & $680$ & $2$ & $0.95$ & $1.00$ \\
12-040 &  DN Tau              & $       2$ & M0        & $  29424$ & 
\hspace*{0.1cm} & $5110$ & $3$ & $0.99$ & $1.00$ & $3072$ & $1$ & $    $ & $
0.97$ & $2040$ & $2$ & $0.98$ & $0.99$ \\
12-059 &  CoKu Tau 3 AB       & $       3$ & M1        & $  29426$ & 
\hspace*{0.1cm} & $10548$ & $1$ & $    $ & $0.98$ & $3547$ & $1$ & $    $ & $
0.56$ & $7000$ & $2$ & $0.98$ & $0.98$ \\
09-022 &  KPNO-Tau 8          & $       3$ & M5.75     & $  29425$ & 
\hspace*{0.1cm} & $1609$ & $3$ & $0.96$ & $0.99$ & $900$ & $1$ & $    $ & $
0.96$ & $721$ & $1$ & $    $ & $0.97$ \\
08-037 &  HQ Tau AB           & $       3$ &           & $  41566$ & 
\hspace*{0.05cm}$^{M1}$ & $3194$ & $2$ & $0.93$ & $1.00$ & $937$ & $1$ & $    
$ & $0.99$ & $2255$ & $2$ & $0.91$ & $1.00$ \\
09-026 &  HQ Tau AB           & $       3$ &           & $  29826$ & 
\hspace*{0.1cm} & $4840$ & $4$ & $0.98$ & $1.00$ & $1519$ & $2$ & $0.98$ & $
1.00$ & $3320$ & $2$ & $0.98$ & $1.00$ \\
08-043 &  KPNO-Tau 15         & $       3$ & M2.75     & $  35322$ & 
\hspace*{0.1cm} & $6014$ & $1$ & $    $ & $0.97$ & $1636$ & $1$ & $    $ & $
0.74$ & $4384$ & $2$ & $0.89$ & $1.00$ \\
09-031 &  KPNO-Tau 15         & $       3$ & M2.75     & $  26257$ & 
\hspace*{0.1cm} & $276$ & $1$ & $    $ & $0.42$ & $140$ & $1$ & $    $ & $0.63
$ & $136$ & $1$ & $    $ & $0.84$ \\
08-048 &  HP Tau AB           & $       2$ & K3        & $  37574$ & 
\hspace*{0.1cm} & $2488$ & $2$ & $0.82$ & $1.00$ & $688$ & $1$ & $    $ & $
0.97$ & $1801$ & $2$ & $0.83$ & $1.00$ \\
08-051a &  HP Tau/G3 AB        & $       3$ & K7        & $  35141$ & 
\hspace*{0.1cm} & $992$ & $1$ & $    $ & $0.87$ & $451$ & $1$ & $    $ & $0.38
$ & $556$ & $1$ & $    $ & $0.27$ \\
08-051 &  HP Tau/G2           & $       3$ & G0        & $  35574$ & 
\hspace*{0.1cm} & $15377$ & $2$ & $0.98$ & $1.00$ & $5871$ & $2$ & $0.98$ & $
1.00$ & $9510$ & $2$ & $0.99$ & $1.00$ \\
08-058 &  Haro 6-28 AB        & $       2$ & M2/M3.5   & $  35457$ & 
\hspace*{0.1cm} & $306$ & $2$ & $0.88$ & $0.19$ & $144$ & $1$ & $    $ & $0.17
$ & $160$ & $1$ & $    $ & $0.59$ \\
08-080 &  CFHT-BD Tau 3       & $       4$ & M7.75     & $  35678$ & 
\hspace*{0.1cm} & $29$ & $1$ & $    $ & $0.99$ & $5$ & $1$ & $    $ & $0.81
$ & $19$ & $1$ & $    $ & $0.99$ \\
05-005 &  CFHT-Tau 6          & $       4$ & M7.25     & $  25463$ & 
\hspace*{0.1cm} & $68$ & $2$ & $0.87$ & $0.97$ & $34$ & $1$ & $    $ & $0.43
$ & $33$ & $1$ & $    $ & $0.99$ \\
05-013 &  GN Tau AB           & $       2$ & M2.5      & $  24825$ & 
\hspace*{0.1cm} & $105$ & $1$ & $    $ & $0.93$ & $13$ & $1$ & $    $ & $0.20
$ & $93$ & $1$ & $    $ & $0.98$ \\
05-017 &  IRAS 04365+2535     & $       1$ &           & $  24037$ & 
\hspace*{0.1cm} & $48$ & $1$ & $    $ & $0.34$ & $-$ & $-$ & $   -$ & $   -
$ & $46$ & $1$ & $    $ & $0.30$ \\
05-024 &  IRAS 04369+2539     & $       2$ & K4        & $  28083$ & 
\hspace*{0.1cm} & $296$ & $3$ & $0.85$ & $1.00$ & $11$ & $1$ & $    $ & $0.94
$ & $286$ & $2$ & $0.83$ & $1.00$ \\
07-011 &  JH 223              & $       3$ & M2        & $  29922$ & 
\hspace*{0.1cm} & $227$ & $1$ & $    $ & $0.69$ & $146$ & $1$ & $    $ & $0.53
$ & $79$ & $1$ & $    $ & $0.52$ \\
07-022 &  Haro 6-32           & $       3$ & M5        & $  29850$ & 
\hspace*{0.1cm} & $157$ & $1$ & $    $ & $0.75$ & $95$ & $1$ & $    $ & $0.16
$ & $62$ & $1$ & $    $ & $0.95$ \\
07-041 &  IRAS 04385+2550AB   & $       2$ & M0        & $  28467$ & 
\hspace*{0.1cm} & $185$ & $3$ & $0.95$ & $0.99$ & $12$ & $1$ & $    $ & $0.93
$ & $175$ & $3$ & $0.96$ & $0.99$ \\
10-017 &  CoKuLk332/G2 AB     & $       3$ & M0.5/2.5  & $  29960$ & 
\hspace*{0.1cm} & $3753$ & $1$ & $    $ & $0.44$ & $1381$ & $1$ & $    $ & $
0.33$ & $2374$ & $1$ & $    $ & $0.56$ \\
10-018 &  CoKuLk332/G1 AB     & $       3$ & K7/M1     & $  30036$ & 
\hspace*{0.1cm} & $420$ & $1$ & $    $ & $0.52$ & $152$ & $1$ & $    $ & $0.67
$ & $268$ & $1$ & $    $ & $0.30$ \\
10-020 &  V955 Tau AB         & $       2$ & K5/M1     & $  31665$ & 
\hspace*{0.05cm}$^{M1}$ & $247$ & $1$ & $    $ & $0.99$ & $28$ & $1$ & $    
$ & $0.74$ & $219$ & $1$ & $    $ & $0.98$ \\
10-034 &  CIDA 7              & $       2$ & M4.75     & $  25405$ & 
\hspace*{0.1cm} & $54$ & $1$ & $    $ & $0.83$ & $40$ & $1$ & $    $ & $0.24
$ & $19$ & $1$ & $    $ & $0.67$ \\
10-045 &  DP Tau              & $       2$ & M0.5      & $  29926$ & 
\hspace*{0.1cm} & $56$ & $1$ & $    $ & $0.98$ & $18$ & $1$ & $    $ & $0.99
$ & $34$ & $1$ & $    $ & $0.99$ \\
10-060 &  GO Tau              & $       2$ & M0        & $  28564$ & 
\hspace*{0.1cm} & $256$ & $1$ & $    $ & $0.78$ & $95$ & $1$ & $    $ & $0.75
$ & $166$ & $1$ & $    $ & $0.60$ \\
\hline
\end{tabular}\end{center}
\end{sidewaystable*}

\addtocounter{table}{-1}

\begin{sidewaystable*}\begin{center}\small
\caption{{\em continued}}
\begin{tabular}{llccrp{0.12cm}rrrrrrrrrrrr}\hline
     &             &      &     &          & & \multicolumn{4}{c}{-------- Broad band --------}                   & \multicolumn{4}{c}{-------- Soft band --------}                    & \multicolumn{4}{c}{-------- Hard band --------}                    \\
XEST No. & Identification & Type & SpT & Expo [s] & & Cts & $N_{\rm b}$ & $CL_{\rm corr}$ & $P_{\rm KS,eff}$ & Cts & $N_{\rm b}$ & $CL_{\rm corr}$ & $P_{\rm KS,eff}$ & Cts & $N_{\rm b}$ & $CL_{\rm corr}$ & $P_{\rm KS,eff}$ \\
\hline\hline
26-012 &  2M J04552333+30     & $       4$ & M6.25     & $ 126315$ & 
\hspace*{0.05cm}$^{M1}$ & $25$ & $1$ & $    $ & $0.65$ & $13$ & $1$ & $    
$ & $0.94$ & $12$ & $1$ & $    $ & $0.57$ \\
26-034 &  2M J04554046+30     & $       3$ & M5.25     & $ 126497$ & 
\hspace*{0.05cm}$^{M1}$ & $19$ & $1$ & $    $ & $0.30$ & $10$ & $1$ & $    
$ & $0.02$ & $10$ & $1$ & $    $ & $0.91$ \\
26-043 &  AB Aur              & $       5$ & B9.5-A0   & $ 128309$ & 
\hspace*{0.05cm}$^{M1}$ & $1404$ & $2$ & $0.98$ & $1.00$ & $954$ & $2$ & $0.98
$ & $1.00$ & $451$ & $1$ & $    $ & $0.99$ \\
26-050 &  2MJ04554757/801     & $       2$ & M4.75/5.6 & $ 126544$ & 
\hspace*{0.05cm}$^{M1}$ & $209$ & $1$ & $    $ & $0.74$ & $118$ & $1$ & $    
$ & $0.67$ & $96$ & $1$ & $    $ & $0.90$ \\
26-067 &  SU Aur              & $       2$ & G2        & $ 128035$ & 
\hspace*{0.05cm}$^{M1}$ & $26452$ & $12$ & $0.98$ & $1.00$ & $6481$ & $5$ & $
0.99$ & $1.00$ & $19971$ & $11$ & $0.98$ & $1.00$ \\
26-072 &  HBC 427             & $       3$ & K7        & $ 128090$ & 
\hspace*{0.05cm}$^{M1}$ & $13791$ & $8$ & $0.98$ & $1.00$ & $6147$ & $6$ & $
0.98$ & $1.00$ & $7643$ & $8$ & $0.97$ & $1.00$ \\
\hline
\end{tabular}\end{center}
\end{sidewaystable*}

\subsection{Results from variability tests}\label{subsect:results_var}

\begin{figure*}
\begin{center}
\parbox{18cm}{
\parbox{9cm}{
\resizebox{9cm}{!}{\includegraphics{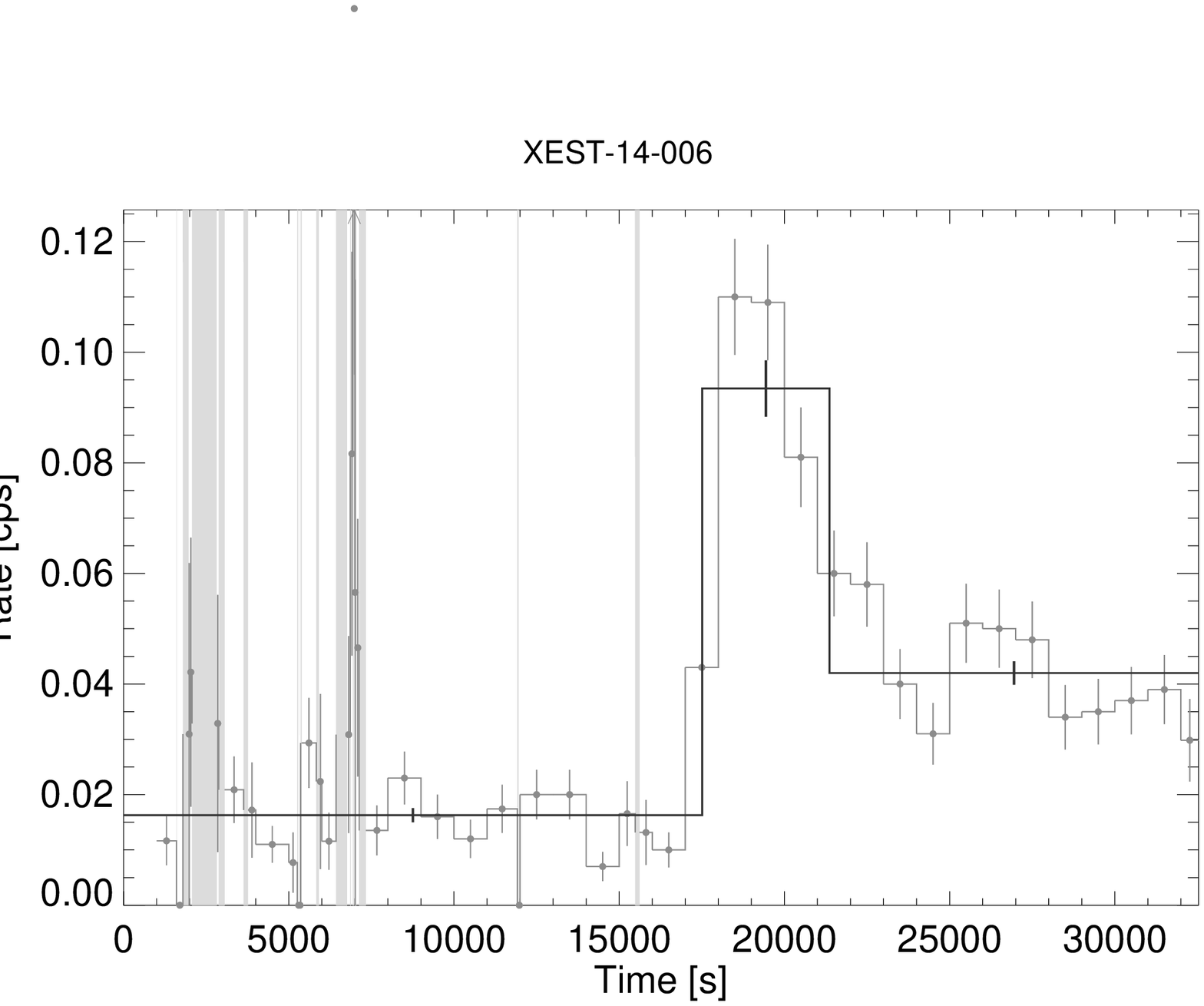}}
}
\parbox{9cm}{
\resizebox{9cm}{!}{\includegraphics{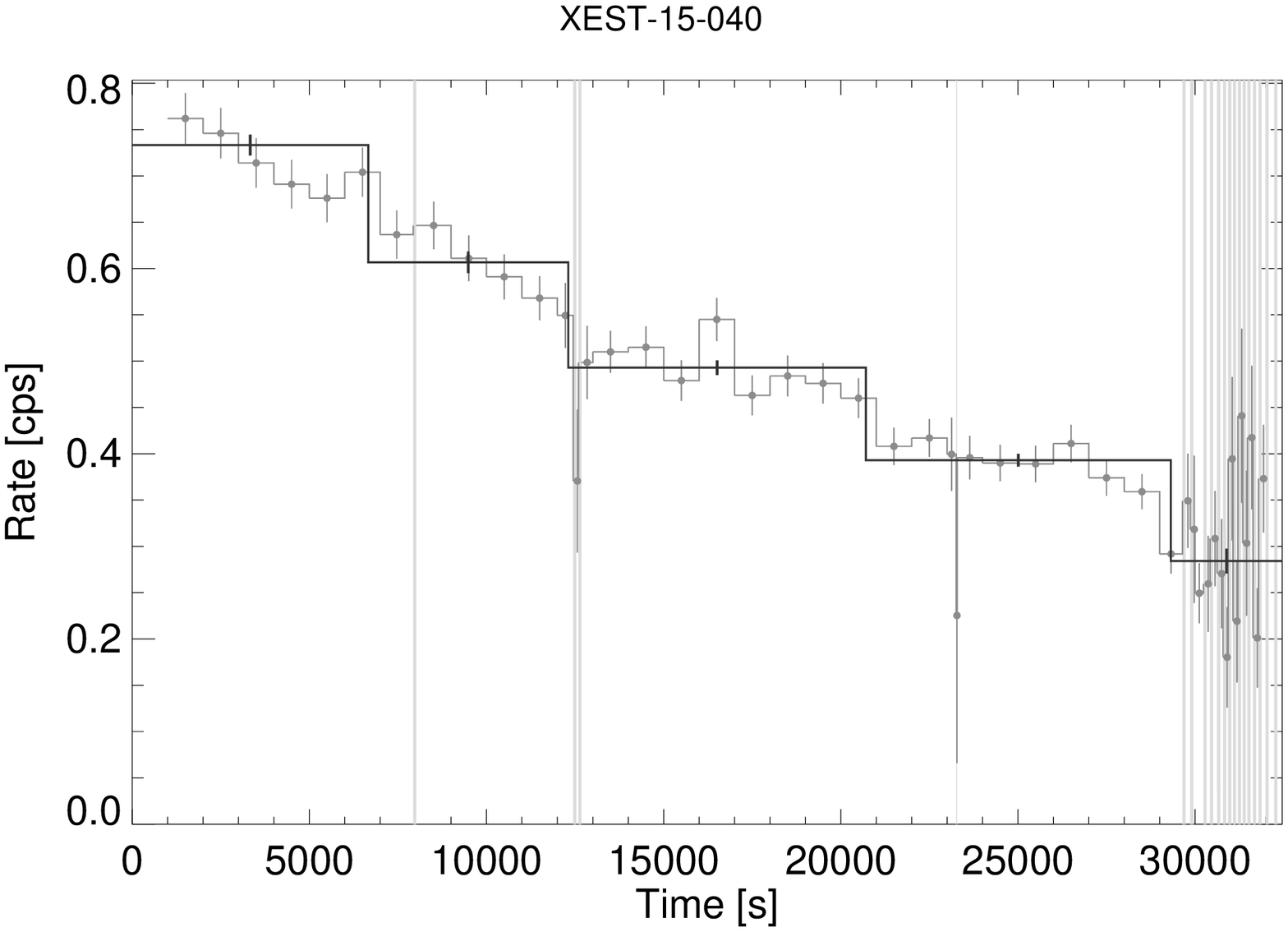}}
}
}
\parbox{18cm}{
\parbox{9cm}{
\resizebox{9cm}{!}{\includegraphics{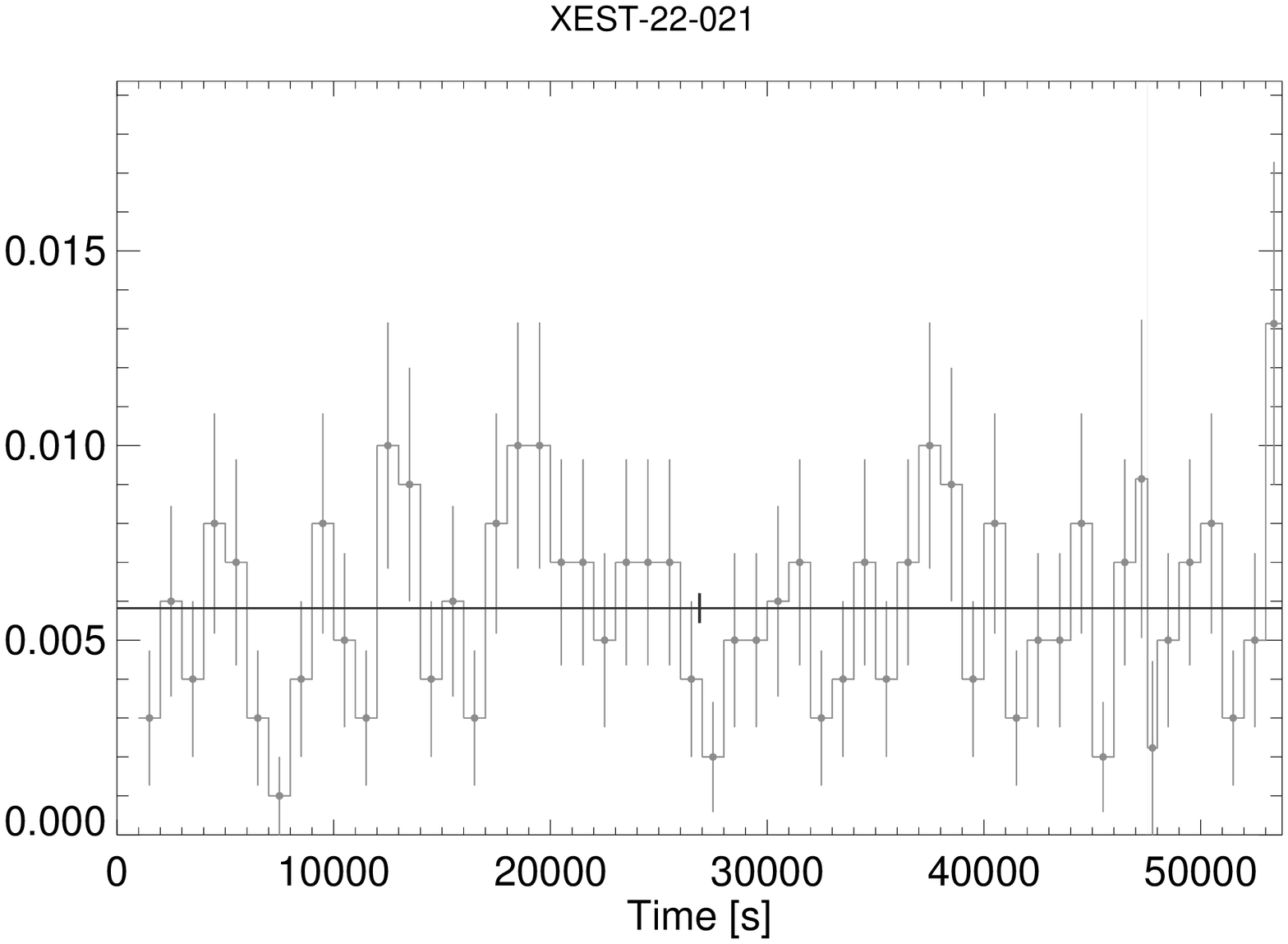}}
}
\parbox{9cm}{
\caption{EPIC/pn time-series for three TMC members, demonstrating the various shapes of the lightcurves: a `canonical' flare (XEST-14-006), a smooth decrease of the count rate (XEST-15-040), and constant signal (XEST-22-021). The blocks resulting from the MLB analysis are overlaid on the binned lightcurve ($1000$\,s bins). Data is background subtracted. Time-intervals removed by the background filter are marked with grey-shades. For clarity the background lightcurve is not shown.}
\label{fig:lcs_diverse}
}
}
\end{center}
\end{figure*}

The variability analysis described in the previous section was applied
to $126$ photon time series from the $122$ different TMC members detected in the XEST. 
The results are given in Table~\ref{tab:var} for all examined sources. 
For a few sources, in the soft and/or hard band the 
variability tests could not be carried out due to poor statistics. 
The entries in Table~\ref{tab:var} are given as in \citet{Guedel06.1} in 
order of increasing right ascension. 
We list the XEST No. (col.~$1$), identification (col.~$2$), 
classification defined by \citet{Guedel06.1} that combines  
the Young Stellar Object class derived from the infrared spectral energy distribution 
and the T Tauri type based on H$\alpha$ equivalent widths (col.~3), 
and the spectral type (col.~$4$). More details concerning the definitions of the stellar parameters
are found in \citet{Guedel06.1}. 
Col.~5 gives the broad band exposure time. Generally, the instrument used is EPIC/pn; the 
exceptions are indicated by labels `M1' for MOS\,1 and `M2' for MOS\,2 at the end of col.~5. 
The remaining columns represent the results of the variability tests. For all 
three energy bands the number of source photons (`Cts'), the number of segments
resulting from the MLB analysis (`$N_{\rm b}$'), the confidence level of the MLB analysis 
(`$CL_{\rm corr}$'),
and the probability for variability according to the KS test ($P_{\rm KS,eff}$) are given; 
see table caption for more details. 

We recall that the parameter describing the variability according to the MLB method ($CL$) 
is a threshold, indicating a lower limit for the significance (or confidence) with which 
the detected variability is physical, i.e. not spurious. The parameter describing the variability according to
the KS test ($P$) is a probability for source variability. Therefore, variability detected with 
a low value for $CL$ is compatible with a high value of $P$. 

In general, 
we find good agreement between the detection of variability with the KS test and with the 
MLB method. 
In this article we distinguish variable from non-variable sources based on the results
from the MLB analysis for the broad band. There are $67$ sources with a number of blocks 
$N_{\rm b} > 1$. 
For most of these sources the KS test probability for variability is very high. Exceptions are
XEST-19-049 and XEST-08-058, with $P_{\rm KS} < 0.95$. After visual inspection of their lightcurves 
we assigned these two sources to the non-variable sample.  
Then, the total number of variable
TMC sources is $65$, representing 65/126 = 52\,\% of all TMC members detected in the XEST.

In Fig.~\ref{fig:lcs_diverse} we display the lightcurves of three TMC members, representing some of the 
typical observed shapes: fast `impulsive' variations on time-scale of hours carrying the signature of flares
(e.g. XEST-14-006), gradual variations possibly representing a fraction of a long-duration event
(e.g. XEST-15-040), and constant emission (e.g. XEST-22-021).

\section{Quantifying variability: Definition of flares}\label{sect:var_quant}

To classify the variability we examined the amplitudes and the timescales involved.
\citet{Wolk05.1} have divided the segments resulting from
a similar MLB analysis in three different types, representing `characteristic', 
`elevated', and `very elevated' intensity. 
The relatively short exposure time of the XEST observations together with the 
typical strong variability of young stars makes it difficult to tell what is the 
characteristic emission level of a given source. Therefore, we follow a somewhat 
simplified scheme, in which we consider the segment with the lowest count rate in a 
given time series as the `characteristic' emission.  
The characteristic count rate is henceforth denoted $R_{\rm ch}$. 

Similar to \citet{Wolk05.1}, for the definition of flares we make use of the 
amplitude and the derivative of the segmented lightcurve. The amplitude is
defined as  
\begin{equation}
A_{\rm i} = \frac{R_{\rm i} - 2\,\sigma_{\rm i}}{R_{\rm ch} + 2\,\sigma_{\rm ch}}
\label{eq:1}
\end{equation}
and the derivative as 
\begin{equation}
\Delta_{\rm i+1} = \frac{R_{\rm i+1} - R_{\rm i}}{\rm MIN[t_{\rm i+1},t_{\rm i}]}\,{\rm [ct\,s^{-2}]}
\label{eq:2}
\end{equation}
In Eq.~\ref{eq:1} and Eq.~\ref{eq:2} $R_{\rm i}$ is the count rate of segment $i$,  
$\sigma_{\rm i}$ its uncertainty, and $t_{\rm i}$ its duration. 

We define as a flare those variability detections for which both of the following two criteria 
are fullfilled: 
(i) the amplitude $A_{\rm i} > 1.5$ in one or more consecutive segments, and 
(i) the maximum of the derivatives in these segments exceeds the threshold 
of $\Delta_{\rm i} > 5\,10^{-5}$.

\subsection{Results from flare detection}\label{subsect:flare_results}

%
%
\begin{table*}\begin{center}
\caption{Parameters of X-ray flares detected on TMC members during the XEST: amplitude $A_{\rm F}$ and derivative $[MAX(\Delta)]_{\rm F}$ are used to identify flares; the duration $\tau_{\rm F}$, characteristic luminosity $L_{\rm ch}$ and flare luminosity $L_{\rm F}$, and the flare energy $E_{\rm F}$ are derived as described in Sect.~\ref{subsect:flare_results}; the last column provides a flag that classifies the shape of the lightcurve \citep[see ][]{Franciosini06.1}.}
\label{tab:flares}
\begin{tabular}{lrrrrrrl}\hline
XEST & $A_{\rm F}$ & $[MAX(\Delta)]_{\rm F}$ & $\tau_{\rm F}$ & $\log{L_{\rm ch}}$ & $\log{L_{\rm F}}$ & $\log{E_{\rm F}}$ & Class \\
No.  &             & [$10^{-5}\,{\rm cts\,s^{-2}}$] & [ks]    & [erg/s]            & [erg/s]           & [erg]             &       \\
\hline\hline
02-022 & $     3.0$ & $    21.5$ & $     9.3$ & $    29.3$ & $    29.5$ & $
    33.5$ &  \\
03-022 & $     2.5$ & $     9.1$ & $    13.5$ & $    29.8$ & $    30.0$ & $
    34.1$ &  \\
04-016 & $     2.8$ & $    27.0$ & $>    11.5$ & $    30.7$ & $    30.8$ & $>
    34.8$ & smooth \\
05-024 & $     4.8$ & $    90.5$ & $>     1.5$ & $    30.3$ & $    30.9$ & $>
    34.1$ &  \\
07-041 & $     3.7$ & $    98.8$ & $     2.7$ & $    29.7$ & $    30.1$ & $
    33.5$ &  \\
09-026 & $     1.9$ & $     6.0$ & $>     4.4$ & $    31.0$ & $    31.0$ & $>
    34.6$ & smooth \\
11-057 & $     6.3$ & $    48.2$ & $>    16.6$ & $    30.4$ & $    30.9$ & $>
    35.1$ & smooth \\
13-004 & $     3.4$ & $    45.0$ & $>     7.2$ & $    29.3$ & $    29.6$ & $>
    33.5$ &  \\
14-006 & $     5.7$ & $   123.0$ & $    15.0$ & $    29.8$ & $    30.2$ & $
    34.4$ &  \\
15-040 & $     2.6$ & $     5.8$ & $>    20.7$ & $    30.8$ & $    30.8$ & $>
    35.2$ & smooth \\
17-058 & $     2.6$ & $     7.2$ & $>     5.5$ & $    29.6$ & $    29.7$ & $>
    33.5$ &  \\
17-066 & $     3.0$ & $    95.5$ & $     9.5$ & $    30.3$ & $    30.6$ & $
    34.6$ & atypical \\
18-004 & $     4.0$ & $    31.1$ & $>     6.0$ & $    29.7$ & $    30.2$ & $>
    34.0$ &  \\
20-005 & $     1.8$ & $     7.1$ & $>     1.2$ & $    30.7$ & $    30.6$ & $>
    33.7$ &  \\
20-022 & $    80.3$ & $  4121.5$ & $     3.1$ & $    30.4$ & $    32.1$ & $
    35.6$ &  \\
20-056 & $     3.6$ & $    58.4$ & $>     7.2$ & $    30.1$ & $    30.5$ & $>
    34.3$ &  \\
22-047 & $     4.0$ & $    12.7$ & $>    41.6$ & $    30.1$ & $    30.4$ & $>
    35.1$ & smooth \\
22-089 & $     2.1$ & $     9.0$ & $>     4.2$ & $    30.3$ & $    30.3$ & $>
    34.0$ & smooth \\
23-002/24-002 & $     5.6$ & $    69.3$ & $    15.2$ & $    29.4$ & $    30.0
$ & $    34.2$ &  \\
23-032/24-028 & $     2.9$ & $    39.5$ & $    46.6$ & $    30.7$ & $    30.7
$ & $    35.4$ & atypical \\
23-033/24-029 & $     8.7$ & $    64.9$ & $     8.6$ & $    29.2$ & $    30.0
$ & $    33.9$ &  \\
23-045/24-038 & $     2.1$ & $   150.5$ & $>    26.9$ & $    29.6$ & $    29.6
$ & $>    34.0$ &  \\
23-047/24-040 & $     1.6$ & $    40.3$ & $    40.9$ & $    30.9$ & $    30.7
$ & $    35.3$ & atypical \\
23-047/24-040 & $     8.2$ & $   174.6$ & $>    17.5$ & $    30.9$ & $    31.6
$ & $>    35.9$ & atypical \\
23-048 & $     7.1$ & $    31.3$ & $>     2.7$ & $    28.7$ & $    29.6$ & $>
    33.0$ &  \\
23-050/24-042 & $    13.0$ & $   637.8$ & $    47.4$ & $    30.0$ & $    30.6
$ & $    35.3$ & impulsive \\
23-063/24-055 & $     3.1$ & $    11.1$ & $    36.0$ & $    29.7$ & $    29.9
$ & $    34.5$ &  \\
23-074/24-061 & $     1.8$ & $     8.5$ & $     6.7$ & $    30.4$ & $    30.3
$ & $    34.2$ & impulsive \\
26-067 & $     3.0$ & $    35.8$ & $    19.9$ & $    31.1$ & $    31.3$ & $
    35.6$ &  \\
26-067 & $     2.0$ & $    23.9$ & $     9.0$ & $    31.1$ & $    31.0$ & $
    35.0$ &  \\
26-067 & $     1.7$ & $    18.3$ & $    13.0$ & $    31.1$ & $    30.9$ & $
    35.0$ &  \\
26-072 & $     5.2$ & $    87.6$ & $    56.0$ & $    30.6$ & $    30.9$ & $
    35.6$ & impulsive \\
28-100 & $     3.2$ & $    62.2$ & $    23.8$ & $    30.2$ & $    30.4$ & $
    34.8$ & impulsive \\
\hline
\end{tabular}\end{center}
\end{table*}

With the procedure described above 
flares are detected on $30$ of the $65$ variable sources. Two sources have shown more
than one event 
(XEST-23-047 with two flares, and XEST-26-067 with three flares). 
The results of the flare detection process are summarized in Table~\ref{tab:flares}. 
For all flaring sources the XEST ID (col.~1),  
the amplitude of the flare $A_{\rm F}$ (col.~2) and the  
maximum of the derivative $[MAX(\Delta)]_{\rm F}$ during the flare (col.~3)
are given. Furthermore, we list the duration $\tau_{\rm F}$ (col.~4), obtained by summing 
the length of all segments that define the flare.
Owing to the short exposure times, only $18$ events are observed in their entirety. 
For the remaining flares the duration given in Table~\ref{tab:flares} is a lower limit. 
In col.~5 the quiescent luminosity $L_{\rm ch}$ is given, 
and cols.~6 and~7 represent the flare luminosity $L_{\rm F}$ and flare energy $E_{\rm F}$. 

Before the conversion to luminosities, the count rates were multiplied with a correction
factor that accounts for the source photons outside the extraction area and for the vignetting. 
Then PIMMS\footnote{The Portable Interactive Multi-Mission Simulator (PIMMS) is accessible at
http://asc.harvard.edu/toolkit/pimms.jsp} has been used to obtain the unabsorbed flux for
a given count rate under the assumption of a 1-T Raymond-Smith model \citep{Raymond77.1} subject to 
photo-absorption. For the column density and temperature of the model we adopted the 
results from the spectral fitting presented in Table~6 of \citet{Guedel06.1} for each
individual X-ray source. In the case of 2-T spectral fits we computed the emission 
measure weighted mean temperature for the input to PIMMS.  
The flux was converted to luminosity assuming a distance of $140$\,pc for all sources. 

This procedure applied to the characteristic rate $R_{\rm ch}$ yields $L_{\rm ch}$,  
while from the average count of all segments that define the flare we obtain the average source 
luminosity during the flare. 
The flare luminosity $L_{\rm F}$ was obtained by subtracting $L_{\rm ch}$ from the average source 
luminosity during the flare. 
Finally, the flare energy was computed by multiplying
$L_{\rm F}$ with $\tau_{\rm F}$. In cases where only the rise or only the
decay were observed, $E_{\rm F}$ is a lower limit to the energy emitted during the event. 

%
%
\begin{table*}
\begin{center}
\caption{XEST X-ray variability statistics for different types of young stars in the broad, soft and hard band derived with the MLB technique. For each energy band the number of detected sources, the number of variable sources, and the number fraction of variables are given; see text for details.}
\label{tab:var_statistics}
\begin{tabular}{lrrrrrrrrrr}\hline
       &      & \multicolumn{3}{c}{Broad band: $0.3-7.8$\,keV} & \multicolumn{3}{c}{Soft band: $0.3-1.0$\,keV} & \multicolumn{3}{c}{Hard band: $1.0-7.8$\,keV} \\
Object & Type & \multicolumn{2}{c}{Number} & Fraction          & \multicolumn{2}{c}{Number} & Fraction         & \multicolumn{2}{c}{Number} & Fraction         \\
type   & identifier & detected & variable & variable & studied & variable & variable & studied & variable & variable\\ \hline
Protostar     & $1$ &   9 &   4 &  44\,\% &  7 &   0 &   0\,\% &   9 &   4 & 
 44\,\% \\
cTTS          & $2$ &  52 &  28 &  54\,\% & 50 &   9 &  18\,\% &  52 &  27 & 
 52\,\% \\
wTTS          & $3$ &  51 &  26 &  51\,\% & 51 &  19 &  37\,\% &  51 &  29 & 
 57\,\% \\
Brown dwarf   & $4$ &   8 &   1 &  13\,\% &  7 &   0 &   0\,\% &   7 &   0 & 
  0\,\% \\
HAeBe         & $5$ &   2 &   2 & 100\,\% &  2 &   2 & 100\,\% &   2 &   1 & 
 50\,\% \\
Unknown       & $9$ &   4 &   4 & 100\,\% &  0 &   0 & $-$     &   0 &   0 & 
$-$     \\ \hline
Total         &     & 126 &  65 &  52\,\%               &117 &  30 &  26
\,\% & 121 &  61 &  50\,\% \\
\hline
\end{tabular}
\end{center}
\end{table*}

The last column of Table~\ref{tab:flares} provides a flag that characterizes the shape of the
lightcurve for the brightest sources with flares, adopted from \citet{Franciosini06.1}.
They present a detailed study of the time-evolution of spectral parameters for the brightest
variable XEST sources, and  
classify them into four groups:
The first group is given by typical or `impulsive' flares with short rise and longer decay. 
Some flares are termed `atypical' because they show a gradual rise. 
The group labeled `smooth' in Table~\ref{tab:flares} is defined by lightcurves that are 
either slowly decaying or slowly rising throughout the whole observation. 
For these sources the characteristic level determined with the MLB procedure is most likely too high.
We did not attempt to classify the flares on the fainter stars. In many cases they are
described by a single elevated block in the MLB analysis, such that their shape is difficult
to determine.

\section{Variability statistics}\label{sect:results_var}

Table~\ref{tab:var_statistics} summarizes the variability statistics for 
different classes of young stars: protostars, cTTS, wTTS, brown dwarfs (BDs), Herbig Ae/Be (HAeBe) stars, and
objects with uncertain classification. The integer numbers for the object type 
identifiers (col.~2) have been introduced by \citet{Guedel06.1}. 
We examined the variability in the soft and hard band, analogously to the broad band. 
Note that in some of the subgroups the sample size (col. labeled `Number studied') 
is slightly smaller in the restricted energy bands due to insufficient statistics. 

In the broad band, variability is detected in about half of the time series. This result 
seems not to depend on the object type. In particular, roughly the same fraction of cTTS
and wTTS are variable. 
For the HAeBe stars the number of objects is too small to draw statistically valid 
conclusions. 
The absence of variability on $7$ of the $8$ detected BDs may be explained by their faintness
that makes it difficult to discern small-amplitude variations (see Sect.~\ref{subsect:flare_bias}). 

The number and the number fraction of variable sources in the hard band are similar to
those in the broad band, but significantly less variations are detected in the soft band
($< 30$\,\% of the total sample are variable at $0.3-1.0$\,keV).
This suggests a tight relation between variability and heating processes.   
On the other hand, in some cases the detection of variability in the soft band may also be impeded by 
extinction. 
%
%
The visual absorption $A_{\rm V}$ is on average 
larger for the sources that show variability in the hard but not in the soft band (`group H'), 
with respect to the sources that are variable at both soft and hard energies (`group H+S'). 
In particular, $\sim 40$\,\% of the sources from group `H' have $A_{\rm V} > 4$\,mag,
compared to only $\sim 10$\,\% of the sources from group `H+S'. 

Table~\ref{tab:flare_statistics} summarizes the flare statistics for the different
types of young stars. 
The flare detection process was run only for the broad band. 
A fraction of $20-30$\,\% of each YSO class with sufficient statistics 
has shown detectable flares. The flare frequency is marginally larger for cTTS with respect to wTTS,
consistent with earlier results based on {\em ROSAT} data \citep{Stelzer00.1}. 
%
%
\begin{table}
\begin{center}
\caption{XEST X-ray flare statistics for different types of young stars in the $0.3-7.8$\,keV broad band derived with the MLB technique. For each YSO class the number of detected sources and the number of flares are given, and the fraction of flares with respect to the total sample.}
\label{tab:flare_statistics}
\begin{tabular}{lcrrr}\hline
Object & Type & \multicolumn{2}{c}{---- Members ----} & Fraction \\
type   & identifier & detected & flare & flare \\ \hline
Protostar     & $1$ &   9 &   2 &  22\,\% \\
cTTS          & $2$ &  52 &  16 &  31\,\% \\
wTTS          & $3$ &  51 &  11 &  22\,\% \\
Brown dwarf   & $4$ &   8 &   0 &   0\,\% \\
HAeBe         & $5$ &   2 &   2 & 100\,\% \\
Unknown       & $9$ &   4 &   2 &  50\,\% \\ \hline
Total         &     & 126 &  33 &  26\,\% \\
\hline
\end{tabular}
\end{center}
\end{table}

Analogous to the different YSO types, we compare the variability and flare statistics 
of stars with different spectral types in Table~\ref{tab:flare_statistics_spt}.  
Only the results for the broad band are shown. The fraction of variable stars drops 
along the spectral type sequence. However, below we show that this is due to an observational
bias rather than a physical effect. 
%
%
\begin{table}
\begin{center}
\caption{XEST X-ray variability statistics for different spectral types in the $0.3-7.8$\,keV broad band; see Tables~\ref{tab:var_statistics} and~\ref{tab:flare_statistics} and text in Sect.~\ref{sect:results_var} for details.}
\label{tab:flare_statistics_spt}
\begin{tabular}{lrrrrr}\hline
Spectral & \multicolumn{3}{c}{------------ Members ------------}   & \multicolumn{2}{c}{----- Fraction -----} \\
type     & detected & variable         & flare                     & variable & flare             \\ \hline
BAF      &   2 &   2 &   2 & 100\,\% & 100\,\% \\
G        &   4 &   4 &   3 & 100\,\% &  75\,\% \\
K        &  35 &  23 &  12 &  66\,\% &  34\,\% \\
M        &  78 &  32 &  14 &  41\,\% &  18\,\% \\
Unknown  &   7 &   4 &   2 &  57\,\% &  29\,\% \\
\hline
Total    & 126 &  65 &  33 &  52\,\% &  26\,\% \\
\hline
\end{tabular}
\end{center}
\end{table}

\subsection{Observational biases}\label{subsect:flare_bias}

Next to residual uncertainties in the background subtraction,
the capability to detect variations is affected by 
Poisson statistics and the inhomogeneous length of the XEST observations. 
The resulting biases are discussed by investigating 
the amplitudes of the blocked lightcurves.

\begin{figure*}
\begin{center}
\parbox{18cm}{
\parbox{9cm}{
\resizebox{9cm}{!}{\includegraphics{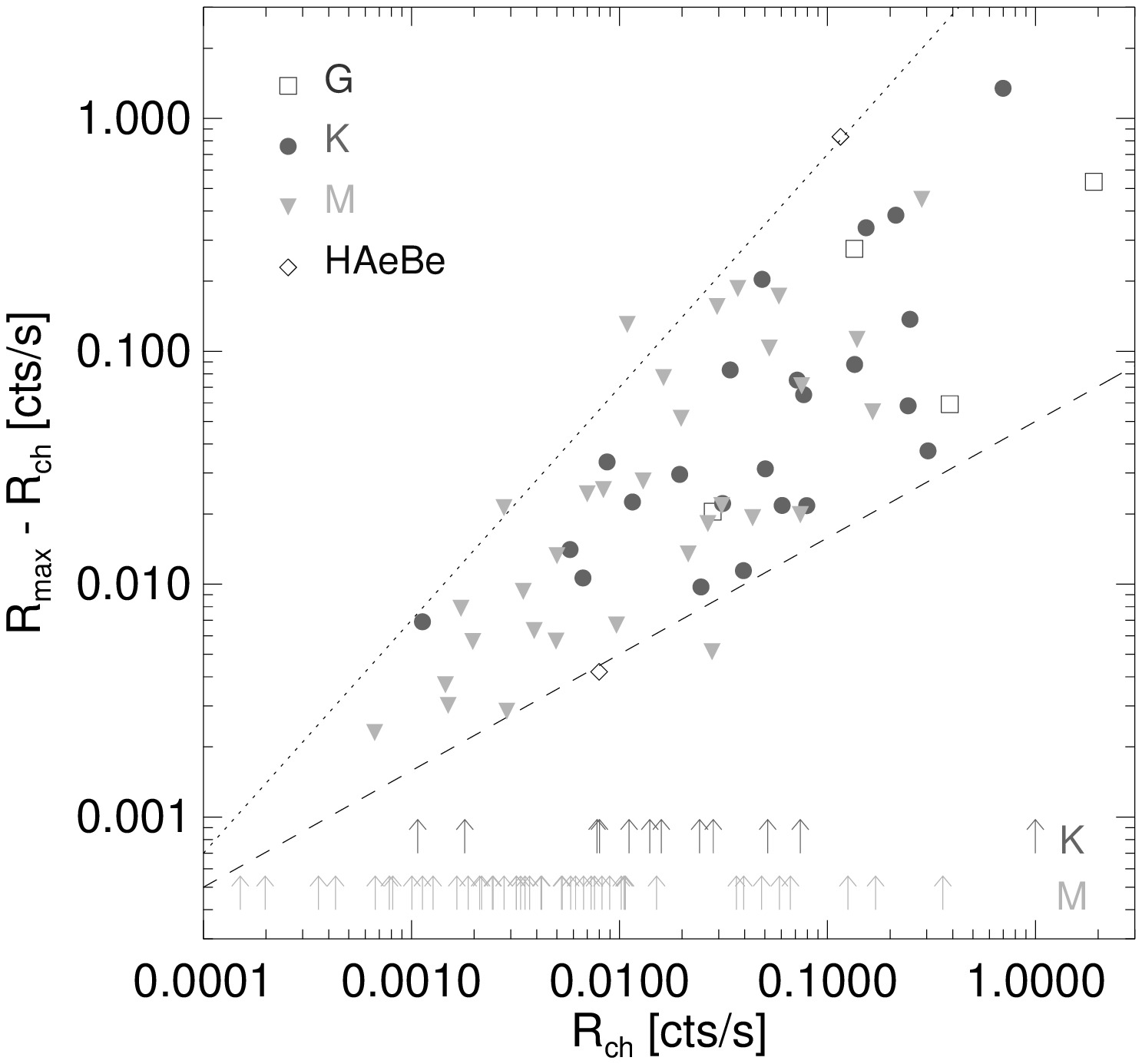}}
}
\parbox{9cm}{
\resizebox{9cm}{!}{\includegraphics{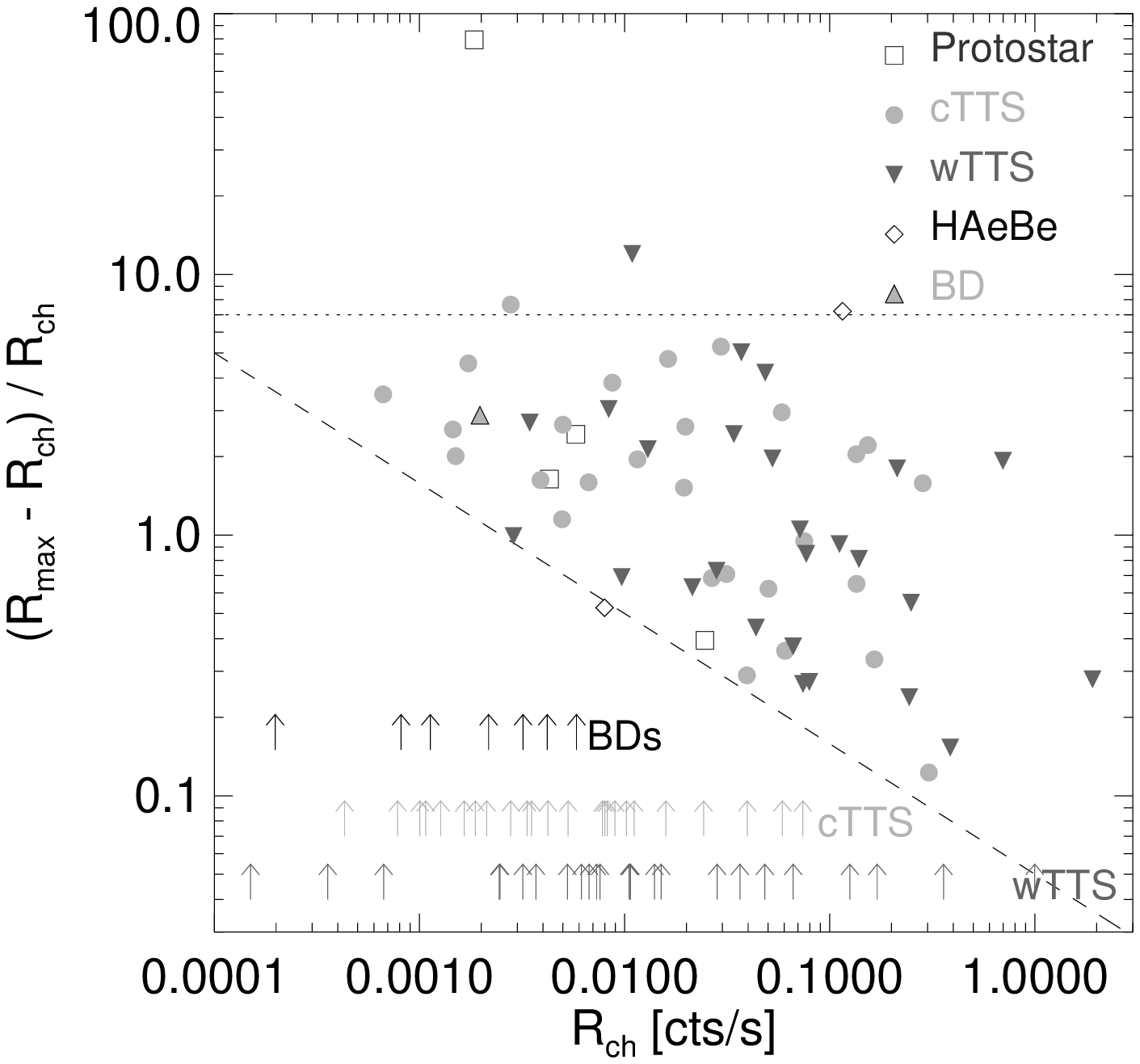}}
}
}
\caption{Absolute amplitude (left) and relative amplitude (right) of the segmented lightcurve 
vs. characteristic count rate for variable stars. On the left the different plotting symbols
distinguish stars with different spectral type, on the right stars of the different YSO classes. 
The dashed line denotes the sensitivity limit according to Poisson statistics
for $\kappa~/\sqrt{t_{\rm max}} = 0.05$. 
The dotted line marks the approximate location of an empirical upper envelope for the 
amplitudes at $R_{\rm max} = 8\,R_{\rm ch}$. Upward pointing arrows are positioned at the average 
count rate of non-variable stars of the different spectral type and YSO groups, respectively.} 
\label{fig:rate_minmax}
\end{center}
\end{figure*}

We define the amplitude of the lightcurves in absolute terms as 
$A_{\rm abs} = R_{\rm max} - R_{\rm ch}$, where $R_{\rm max}$ is the highest block count
rate. 
The left diagram in Fig.~\ref{fig:rate_minmax} 
shows $A_{\rm abs}$ of all variable
stars with known spectral type as a function of the characteristic rate. 
Clearly, the absolute amplitude is correlated with $R_{\rm ch}$.
The existence of a lower envelope is due to the statistics that impose a sensitivity limit 
on the detection of variability. In order to be recognized as variable, the signal $S_{\rm i}$ in a
given block must exceed a certain signal-to-noise level, $\kappa$: 
\begin{equation}
S_{\rm i} = N_{\rm i} - N_{\rm ch,i} > \kappa \sigma(N_{\rm ch,i}) = \kappa \sqrt{N_{\rm ch,i}}
\label{eq:sn}
\end{equation}   
where $N_{\rm i}$ is the observed number of counts in block $i$, $N_{\rm ch,i}$ is the number of 
counts in block $i$ that represent the characteristic level,
and $\sigma$ is the standard deviation as a measure for the noise. Transformed to count rates, this
yields for the absolute amplitude
\begin{equation}
A_{\rm abs} = \frac{1}{t_{\rm max}} \cdot (N_{\rm max} - N_{\rm ch,max}) > \kappa \sqrt{\frac{1}{t_{\rm max}}} \cdot \sqrt{R_{\rm ch}}
\label{eq:abs_ampl}
\end{equation}
This sensitivity limit is approximated by the dashed line in 
Fig.~\ref{fig:rate_minmax}. 
Its location
is not uniquely determined, because it depends on the duration of the block with the maximum count rate. 
For reasonable values of
$t_{\rm max} \sim 5...20$\,ksec the plotted line corresponds to $\kappa \sim 3.5....7$.   
%
%

On the right hand side of Fig.~\ref{fig:rate_minmax} 
the relative amplitude, $A_{\rm rel} = (R_{\rm max}-R_{\rm ch})/R_{\rm ch}$ is shown. 
In this representation the sensitivity threshold has a functional dependence of $1/\sqrt{R_{\rm ch}}$. 
Therefore, events that are small in relative terms are more easily detected on bright stars,
while events that are small in absolute terms are more easily detected on faint stars. 

An approximate empirical upper envelope to the observed variations is marked in 
Fig.~\ref{fig:rate_minmax} 
with a dotted line.
In contrast to the lower threshold that is -- as explained above -- induced by our sensitivity limit,
the absence of large amplitudes is probably the result of the intrinsically rare occurrence of 
strong intensity changes combined with our limited observing time. 
During the XEST the majority of TMC members did not undergo variations by more than a factor 
$8$ in count rate. 

\begin{figure}
\begin{center}
\resizebox{9cm}{!}{\includegraphics{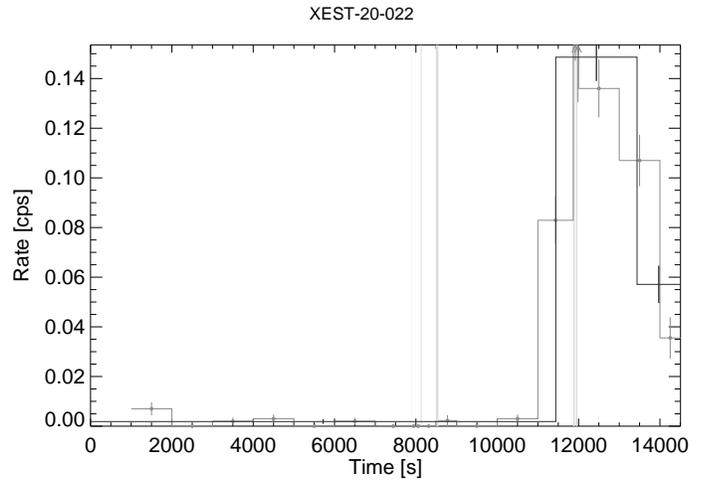}}
\caption{Background subtracted broad band lightcurve of the protostar IRAS-04108+2803\,B (XEST-20-022) 
showing a large flare. The data is dominated by the hard band. The soft emission is absorbed due to high
column density.} 
\label{fig:lc_xest-20-022}
\end{center}
\end{figure}

The different plotting symbols in the right diagram of Fig.~\ref{fig:rate_minmax} distinguish not 
spectral types but the YSO classes from Table~\ref{tab:var_statistics}. 
An exceptionally large amplitude is shown by \iras~(XEST-20-022), a Class\,I protostar. 
This object has shown a spectacular flare near the end of the observation, 
with a rise in count rate by a factor of $\sim 80$. 
Its broad band lightcurve is displayed in 
Fig.~\ref{fig:lc_xest-20-022}. 
The pre-flare quiescent count rate of \iras~ is $\sim 1.9\,10^{-3}$\,cps in EPIC/pn,
near the detection limit of the observation. 
According to the spectral analysis carried out by \citet{Guedel06.1} the source suffers from strong extinction 
($\log{N_{\rm H}}{\rm [cm^{-2}]} \sim 22.9$). Indeed, the soft emission is completely absorbed, 
explaining the sharp contrast. 
%
%

Combining the statistical sensitivity threshold with the empirical maximum of the observed amplitudes,  
the dynamical range for detectable variations is much larger for bright stars than for faint stars.
Therefore, one expects to find more variables among stars with high characteristic emission. To demonstrate
the distribution of $R_{\rm ch}$ in the total sample, at the bottom
of 
Fig.~\ref{fig:rate_minmax} 
the positions of non-variable stars of the different subgroups are indicated. 
For $R_{\rm ch} > 0.3\,{\rm ct\,s^{-1}}$ about $50$\,\% of the sample is variable. 
Similar observations have been discussed in the literature, e.g. \citet{Stelzer05.1}, \citet{Flaccomio06.1}
have shown that for a sample in the field of a given X-ray observation all stars 
above a certain number of counts are variable. 
Probably longer observations would yield the same result for the TMC sample.

\subsection{Variability on different groups of pre-MS stars}\label{subsect:var_groups}

Taking account of the biases discussed above, we can proceed to a further 
examination of the variability on different subsamples of TMC members. 
A comparison of Table~\ref{tab:flare_statistics_spt} with 
Fig.~\ref{fig:rate_minmax} (left) 
suggests that the smaller fraction of variables among M stars
when compared to K stars is a result of their fainter characteristic emission. 
Indeed, the vast majority of M stars gather at $R_{\rm ch} < 0.01\,{\rm ct\,s^{-1}}$ 
, where the detection of small variations is impeded by Poisson statistics. 
The K stars follow a wider distribution including higher values of $R_{\rm ch}$,
such that the detection of variability is favored. 
%
%
When only sources with $R_{\rm ch} < 0.02\,{\rm ct\,s^{-1}}$ 
are considered, the fraction of variable sources is $46 \pm 19$\,\% for the K stars, 
and the fraction of variables is $50 \pm 12$\,\% for the M stars, 
i.e. the variability statistics for the two spectral type classes are indistinguishable.  

In an analogous way, the variability statistics of cTTS and wTTS can be compared. 
It was repeatedly shown in the literature \citep{Neuhaeuser95.1, Stelzer01.1} that 
wTTS are on average X-ray brighter than CTTS. This has also been confirmed for the XEST 
\citep{Guedel06.1}, and therefore one may expect to find a reduced fraction of variable
sources among cTTS with respect to wTTS. 
However, from Table~\ref{tab:var_statistics} it results that a similar fraction of cTTS and wTTS ($\sim 50$\,\%) 
is variable. 
%
%
As seen from 
Fig.~\ref{fig:rate_minmax} (right) 
both sub-groups have a large
fraction of faint objects on which variability is difficult to detect 
($\sim 65$\,\% of all cTTS and $45$\,\% of all wTTS have $R_{\rm ch} < 0.02\,{\rm ct\,s^{-1}}$). 
In the faint subsample with $R_{\rm ch} < 0.02\,{\rm ct\,s^{-1}}$ the fraction of variable sources among the cTTS 
is $41 \pm 11$\,\%, and the fraction of variables among the wTTS is $26 \pm 11$\,\%. This is a significant
reduction with respect to the sample without restriction in $R_{\rm ch}$, but the results for 
cTTS and wTTS are undistinguishable within the statistical uncertainties. 

\begin{figure}[t]
\begin{center}
\resizebox{8cm}{!}{\includegraphics{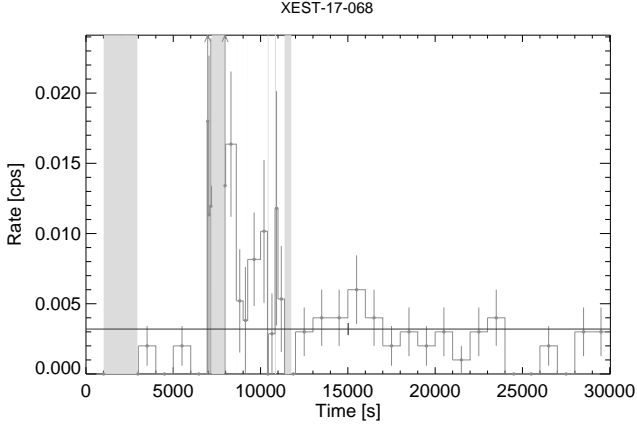}}
\caption{Background subtracted broad band EPIC/pn lightcurve of the BD CFHT-BD-Tau\,1 ( = XEST-17-068). There is some evidence for the possible flare reported by \protect\citet{Grosso06.1} at $t \sim 7000-10000$\,s.}
\label{fig:lc_xest-17-068}
\end{center}
\end{figure}

From 
Fig.~\ref{fig:rate_minmax} 
it can also be seen that the non-detection of variability
on most of the BDs is likely to be related to their low characteristic count rates:  
The XEST is sensitive only to large, and presumably rare, flares on the BDs. 
\citet{Grosso06.1} have identified a flare in the binned lightcurve of 
CFHT-BD-Tau\,1 (XEST-17-068). Our MLB algorithm does not recover this event, 
although there is some evidence for variability in the binned lightcurve. 
According to the KS test the broad band photon time series of XEST-17-068 is variable 
at the $99$\,\% confidence level. 
Possibly this flare was missed in the MLB analysis due to high background.  
Its broad band lightcurve, displayed in Fig.~\ref{fig:lc_xest-17-068}, shows that
part of the flare was removed from the data with our high-background filter.  

\section{Flare duration and frequency}\label{sect:flare_tau}

The duration of the observed flares ranges from a few ksec to $\sim60$\,ksec,
with a clustering of events near $\sim 10$\,ksec. 
Obviously, the XEST is biased
against the detection of long-duration events, due to the short
exposure times (typically $\sim 30$\,ksec). Recall that the distribution of
flare durations derived from the $13$\,d-long COUP exposure \citep{Wolk05.1} 
ranges from $1$\,hr to $3$\,d with a peak near $18$\,h. 
Consequently -- leaving apart subtle differences in the definition of a flare
in the work by \citet{Wolk05.1} and the one presented here, 
and ignoring that the two studies concern stars of different mass range
that may influence the flare characteristics 
-- it is no surprise that 
nearly half of the flares we observed in the XEST are only partly within the 
observing time. 
As mentioned above we can give only lower limits on the duration of these events. 

The frequency of flares for the total XEST sample of TMC members 
computed from the total observing time $T_{\rm obs}$ and the number of observed flares 
is $1$ flare per star in $200$\,ksec. 
In computing
this number we have assumed that we are sensitive to the detection of flares 
throughout the whole observing time. This should be a reasonable approximation, because 
observing gaps due to high background are mostly shorter than the typical duration of 
the flares. 
A serious limitation, however, comes from the fact that we are comparing stars whose 
characteristic count rates differ by as much as a factor of $1000$. 
As a consequence the events detected in our observations form an
incomplete subsample of the total flare population. 
In Sect.~\ref{sect:flare_energy} we argue that the flare detection is complete
for events with $E_{\rm F} > 10^{35}$\,erg/s. If only those events are considered 
the flare frequency of the TMC members reduces to $1$ flare in $770$\,ksec. 
Our flare detection is also biased by the limited length of the XEST observations, typically
$\sim 30$\,ksec. Indeed, during the long ($115$\,ksec) merged observation XEST-23/XEST-24 $8$ flares were
identified on the $18$ stars detected in both exposures XEST-23 and XEST-24. 
Counting only the events with energy above $10^{35}$\,erg the XEST-23/XEST-24 field 
has a flare frequency of $1$ in $520$\,ksec. 

These numbers can be compared to the flare frequency observed during the COUP for the young solar-analogs 
in the ONC. There are $30$ flares that emit more than $10^{35}$\,erg/s \citep[Table~6 of][]{Wolk05.1}, 
and -- following the discussion in the COUP article -- we estimate that their occurrence rate is 
$\sim 1/900$\,ksec, similar to our result for a sample of wider mass range in the TMC. 

The fraction of the total observing time during which a given group of stars was found in the
flare state is defined by 
\begin{equation}
F_{\rm F} = \frac{\sum_i \tau_{\rm i}}{T_{\rm obs}}
\end{equation}
where the sum goes over the durations of all flares. 
For the total TMC sample we find $F_{\rm F} \sim 8$\,\%. 
For the subgroups of cTTS and wTTS the fractional time in the flare state is 
found indistinguishable in both cases at $\sim 8-9$\,\%. 
\citet{Stelzer00.1} have found $\sim 1$\,\% for the flare
rate of TMC stars, evaluating pointed {\em ROSAT} observations. 
This discrepancy may indicate that {\em ROSAT} missed a substantial number of flares
due to gaps in the observing sequence resulting from Earth occultations, or due to uncertainties
in the flare durations related to these gaps, or due to the smaller effective area or 
missing sensitivity for more energetic photons.

\section{Flares and the coronal heating process}\label{sect:flare_energy}

To study the effect of the flare population on coronal heating, 
we now construct the distribution of flare occurrence rate
in total released energy. For the Sun, it was shown that this distribution follows a
power law \citep[e.g.][]{Hudson91.1}
\begin{equation}
\frac{dN}{dE} \sim E^{-\alpha}
\label{eq:flare_energ}
\end{equation}
The radiated energy released in flares in the observed energy band 
is obtained by integrating this differential distribution. Generally, 
observational studies are affected by a sensitivity limit that impedes the detection of
very small flares. It is common practice to extrapolate the power-law to energies 
below the detection threshold in order to get hold of the energy in small unresolved
events. Such `nano-flares' are assumed to exist because they can be 
considered as the heating agent that 
gives rise to the quiescent corona of the Sun and other magnetically active stars. 
If the power-law index $\alpha > 2$, very small flares can in principle contribute an 
unlimited amount of energy, because the integral of Eq.~\ref{eq:flare_energ}
diverges for $E_{\rm min} \rightarrow 0$. 
Clearly, in such cases there must either be a lower cut-off to the
flare energies, or the power law must turn over to $\alpha < 2$ at low
energies. 

Although we are limited to measurements exceeding $\log E_{\rm F}\,{\rm [erg]} \approx 33.5$
in our survey, we note that 
a power-law distribution has been followed down to luminosities of $ \log E_{\rm F}\,{\rm [erg]} = 30.5$
for a few examples of somewhat more evolved, magnetically active low-mass stars 
and to levels as low as $ \log E_{\rm F}\,{\rm [erg]} = 25$ in the solar corona.
For the Sun the situation has remained inconclusive, with $\alpha$ between $1.6-2.6$
over a wide range of energies \citep{Crosby93.1, Krucker98.1, Aschwanden00.1}. 
For stars the hard X-ray range, mostly used in solar flare observations, is inaccessible. 
In recent statistical studies of EUV flare energy distributions on magnetically 
active stars \citet{Audard99.1} and \citet{Audard00.1} find power-law indices
close to the critical value ($\alpha \sim 2$). 
Further observations of stellar flare energy distributions are summarized by 
\citet{Kashyap02.1} and \citet{Guedel03.1}, who point out that 
the slope may depend on the spectral range of the instrument. 
A comparison of the distributions observed in the EUV, in soft, and in hard X-rays
showed that these different temperature regimes correspond to different energy ranges
from `nano'- to `milli-flares', and that they represent different density, 
pressure, and emission measure regimes \citep{Aschwanden00.1}. 
Therefore, it is unclear if the extrapolation from the large observable flares 
towards the lowest energies is valid. 

The analysis of such distributions is subject to several caveats. One of them is
that small flares overlapping larger ones are masked and missed in the flare 
detection process. 
\citet{Audard00.1} have mediated this problem by 
applying a correction factor to the flare rate at each energy. 
This correction takes account of the fact that 
the `effective' time available for the identification
of small flares is given by the total observing time ($T_{\rm obs}$) 
reduced by the sum of the durations of larger flares ($\tau_{\rm > E}$), such that the
flare rate at energy $E$ must be multiplied by a factor  
\begin{equation}
f_{\rm E} = T_{\rm obs} / ( T_{\rm obs} - \tau_{\rm >E} ). 
\label{eq:flare_rate_correction}
\end{equation}

\subsection{Flare energy distributions: TMC and ONC}\label{subsect:energ_distr}
                                                   
\begin{figure}
\begin{center}
\resizebox{8cm}{!}{\includegraphics{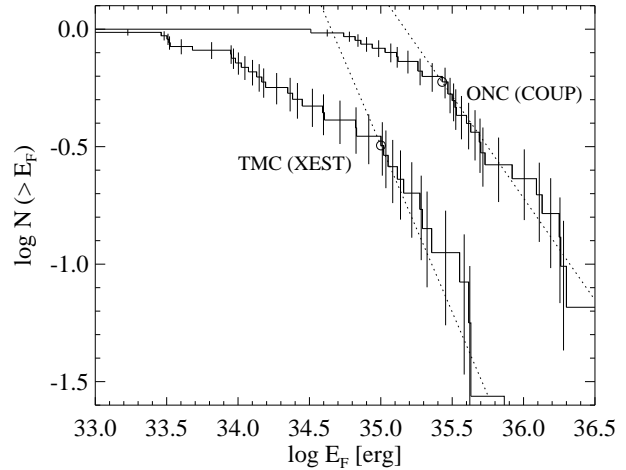}}
\caption{Cumulative distribution of flare energies for the pre-MS sample in the TMC observed during the XEST (Table~\ref{tab:flares}) and for the solar-analogs of the ONC observed during the COUP \protect\citep[Table~6 of][]{Wolk05.1}.  
The dotted lines represent the ML estimates for the slope of the distribution $\log{N} = \beta \cdot \log{E} + C$ in the range of energies above the cutoff value $E_{\rm cut}$ indicated by the open circle.}
\label{fig:cumdist_ef}
\end{center}
\end{figure}

Systematic studies of the energy release during flares on pre-MS stars have
been carried out for only two star forming regions so far: the TMC (this study)
and the solar-analogs of the ONC \citep{Wolk05.1}. In Fig.~\ref{fig:cumdist_ef} 
the cumulative number distribution [$N(>E_{\rm F})$] of flares observed in both
samples are shown  
in double logarithmic form. 

We derived the distribution of the ONC from the mean flare energies 
given in Table~6 of \citet{Wolk05.1}. In contrast to our XEST study of the TMC, 
these energies were computed from the spectrum during the flare state. 
In some cases the values for $E_{\rm F}$ are dubious as a result of poor spectral fits, and we 
excluded events flagged with `c', `e', or `f' in Table~6 of \citet{Wolk05.1} from the analysis
of the energy distribution. 

We performed for both distributions the correction for overlapping flares
described above by multiplying each value of $N(>E)$ with its appropriate factor $f_{\rm E}$. 
For the ONC data we assumed 
a total observing time of $850\,{\rm ksec} \cdot 22$, where $22$ is
the number of stars in the sample after removing the $5$ stars with poorly defined 
flares mentioned above. 
The duration of each flare is taken from Table~6 of \citet{Wolk05.1}. 
The effect of the correction factor $f_{\rm E}$ is to steepen the slope, 
but in practice its influence is negligible because the flare frequency is low, such
that $\tau_{\rm >E}$ is only a minor fraction of $T_{\rm obs}$. 

The faintest flare detected in the ONC sample has an energy of $10^{34.5}$\,erg/s, 
while in the TMC we have access to events that are one order of magnitude lower. 
This difference in sensitivity corresponds roughly to the difference in X-ray flux for a 
source of given luminosity that arises from the distance ratio $(d_{\rm ONC}/d_{\rm TMC})^2$ 
($140$\,pc for TMC vs. $450$\,pc for the ONC).
%
%

\subsection{Determination of the power-law slope}\label{subsect:energ_slope}

As discussed above, the distribution of flare energies is expected to follow a power-law. 
However, the observed distributions flatten towards lower energies.  
This saturation is 
probably an observational bias, indicating that smaller events are missed by the 
flare detection process. On the other hand, the steep high-energy portion represents large flares 
that are readily detectable, such that above 
a certain threshold energy $E_{\rm cut}$ 
the flare identification is likely complete. 

We have determined both the slope $\beta$ of the function 
$\log N(>E_{\rm F}) = - \beta \cdot \log{E_{\rm F}} + C$ and the cutoff energy $E_{\rm cut}$ with a 
maximum likelihood method. According to \citet{Crawford70.1} the ML estimate for the power law slope 
is 
\begin{equation}
\frac{1}{\beta} = \frac{1}{N} \sum_i{\ln{(\frac{E_{\rm i}}{E_{\rm cut}})}}
\label{eq:beta}
\end{equation}
where $N$ is the total number of flares above $E_{\rm cut}$, and 
$E_{\rm i}$ are their energies. To test if Eq.~\ref{eq:beta} is a good representation of the data 
we have performed the transformation 
\begin{equation}
y_{\rm i} = ( 1 - (\frac{E_{\rm i}}{E_{\rm cut}})^{-\beta} ) / (1 - (\frac{E_{\rm max}}{E_{\rm cut}})^{-\beta} )
\label{eq:yps}
\end{equation}
and tested this distribution against departure from a uniform distribution using the KS test
\citep{Crawford70.1}.  

\begin{figure}
\begin{center}
\resizebox{8cm}{!}{\includegraphics{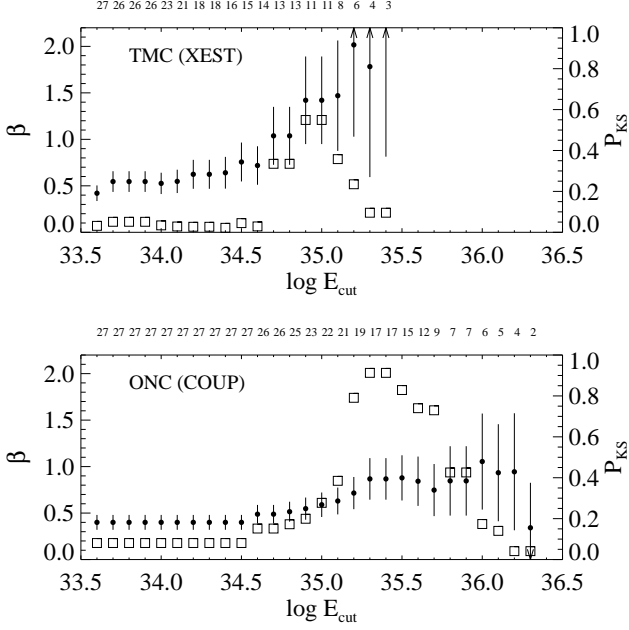}}
\caption{Result from the ML estimate of the power-law index for the flare energy distribution of the XEST sample in the TMC and the COUP sample in the ONC as a function of minimum flare energy included ($E_{\rm cut}$): {\em filled circles} - slope $\beta$ of the double logarithmic cumulative distribution of flare energies ($\log{N} = - \beta \cdot \log{E} + C$); {\em open squares} - probability of KS test that the above power law is a good approximation of the observed flare energy distribution.}
\label{fig:beta}
\end{center}
\end{figure}

We have evaluated Eq.~\ref{eq:beta} and~\ref{eq:yps} for a range of cutoff energies. 
The result is shown in Fig.~\ref{fig:beta}. It presents in the same graph, but with different
scales, the run of $\beta$ (filled circles) and the probability $P_{\rm KS}$ from the KS test 
(open squares).
High values of $P_{\rm KS}$ indicate that the distribution of $y_{\rm i}$ is compatible with
the null hypothesis of being uniform in the interval [0,1]. Therefore, the power law approximation
of the observed flare energy distribution is justified for cutoff energies in 
the peak of $P_{\rm KS}\,{\rm (E_{\rm cut})}$. From Fig.~\ref{fig:cumdist_ef} the power law index 
$\beta$ is expected to increase with increasing $E_{\rm cut}$. For cutoff energies where
the flare energy distribution is well represented by a power law (i.e. for high values of $P_{\rm KS}$), 
the flare detection process is probably complete. 
Therefore, further increasing $E_{\rm cut}$ should not change the slope, and
$\beta\,{\rm (E_{\rm cut})}$ reaches a plateau.  
Obviously, at the largest cutoff energies the result is compromised by low-number statistics. 
The number of flares with $E_{\rm F} > E_{\rm cut}$ are given on the top of both panels in 
Fig.~\ref{fig:beta}. 
%
%

\subsection{Results and caveats}\label{subsect:energ_results}

The best match for the power-law index determined from Fig.~\ref{fig:beta} is 
$\beta_{\rm TMC} = 1.4 \pm 0.5$ for the TMC and $\beta_{\rm ONC} = 0.9 \pm 0.2$ for the ONC. 
These slopes are plotted in Fig.~\ref{fig:cumdist_ef}, and the corresponding $E_{\rm cut}$
is marked with an open circle. It suggests that the flare detection is complete for 
events of $\log{E_{\rm F}}\,{\rm [erg]} > 34.9$ in the case of the TMC, 
and $\log{E_{\rm F}}\,{\rm [erg]} > 35.3$ for the ONC. 
The cumulative distribution of Fig.~\ref{fig:cumdist_ef} 
has the derivative $dN/dE_{\rm F} \sim E_{\rm F}^{-\alpha}$, 
where $\alpha = 1 + \beta$. Therefore, from our observations we find 
$\alpha_{\rm TMC} \sim 2.4$ and $\alpha_{\rm ONC} \sim 1.9$. Both values are consistent
with each other and with $\alpha > 2$, considering the substantial uncertainties. 

Besides the statistical errors, there are various uncertainties 
and differences in the approach between the ONC and the TMC sample, 
e.g. the definition of the flares and the evaluation of the flare energies. 
In our analysis we have ignored that the observed energies for some flares in the XEST 
are lower limits. No information on a possible incomplete coverage of the ONC 
flares is given by \citet{Wolk05.1}. However, as judging from their Fig.~3, some of the events 
should be flagged lower limits because data gaps during the flare 
have caused a loss of photons for the spectral analysis. 
We have verified in simulations that the power-law index $\beta$ is not affected by ignoring
the censoring of some energies. 
If a certain number of values at random drawn from a power-law distribution are 
decreased by a random fraction of their original value, 
this results in a leftward shift of the cumulative number distribution, but
the slope remains unmodified. This means that we implicitly assume a random distribution
of the lower limits across the range of observed flare energies. While in practice this assumption
remains unproven, this is unlikely to be of major importance to our results.  

We also note that we did not perform a thermal modeling of
individual flares to estimate their radiated energies in detail,
given that the spectra were too faint in many cases to obtain
reliable temperature, emission measure, and absorption parameters.
It was implicitly assumed that the flare parameters are similar
to the coronal parameters determined for the overall emission.
Given the relatively moderate dynamic range of flare amplitudes
and also their moderate amplitudes compared with the overall
emission level in the respective light curves,
we believe that this approximation is appropriate.

\subsection{Completeness limit for flare detection}\label{subsect:energ_sens}

We estimate the minimum signal for the detection of a flare with edge 
energy $E_{\rm F,edge} = 10^{35}$\,erg  
and a typical duration of $10$\,ksec. 
For such events the flare luminosity is 
$\log{L_{\rm F,edge}}{\rm [erg/s]} = 31$. 
We compute the value of $L_{\rm F,edge}/L_{\rm ch}$ that corresponds 
to this threshold from the characteristic count rate of each star. 
Again, we use PIMMS with spectral parameters from Table~6 of \citet{Guedel06.1} to determine
an individual count-to-flux conversion factor for each source, that allows to determine $L_{\rm ch}$. 

\begin{figure}
\begin{center}
\resizebox{8cm}{!}{\includegraphics{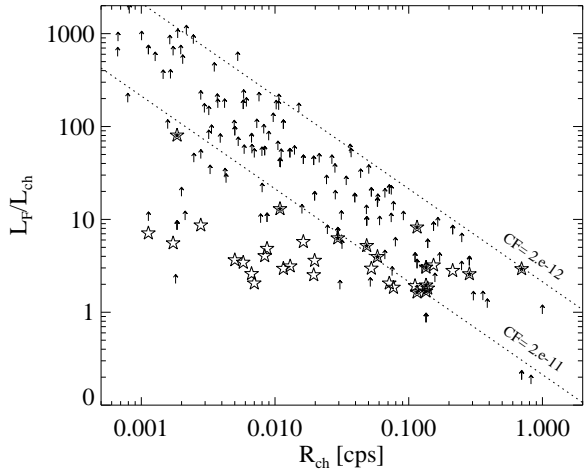}}
\caption{Flare sensitivity limit. Shown is for each star the amplitude above which all flares with an 
assumed emitted energy of $10^{35}$\,erg/s and a duration of $10$\,ksec are detectable (upward pointing arrows). 
The conversion from count rate to luminosity depends on the spectral shape, and is expressed by the dotted 
lines that indicate two different count-to-energy conversion factors. 
Asterisks denote the amplitudes of observed flares; filled asterisks are those 
flares with $E > 10^{35}$\,erg/s, and the flare sensitivity limit of the corresponding
star is shown as thicker arrow.} 
\label{fig:flare_completeness}
\end{center}
\end{figure}

The resulting amplitude $L_{\rm F,edge}/L_{\rm ch}$ 
is visualized in Fig.~\ref{fig:flare_completeness}. The choice of upward pointing arrows as 
plotting symbol is motivated by the indications discussed in Sect.~\ref{subsect:energ_slope}  
that all flares with energy above the cutoff 
$E_{\rm F,edge}$ are detected, such that $L_{\rm F,edge}/L_{\rm ch}$ represents a completeness threshold.  
As can be seen, 
for given $R_{\rm ch}$ the differences in the spectral shape result in a spread of more than one 
order of magnitude in the luminosity amplitude. 
This is illustrated by the dotted lines that represent two different conversion factors. 
For comparison in Fig.~\ref{fig:flare_completeness} the amplitudes of all observed flares are also shown 
(asterisks). 
Generally, the `completeness limit' corresponds to much higher amplitudes than observed, leaving
space for a large population of undetected events. 
Filled asterisks denote those flares with $E_{\rm F} > 10^{35}$\,erg. 
As expected most of these are well above the derived completeness limit for their
host star (marked with thicker arrows). 
In some cases they are below, and this can be explained by their duration being much longer
than our assumption of $10$\,ksec. 
%
%

\section{Variability of cTTS and wTTS}\label{sect:ctts_wtts}

A comparison of the variability characteristics of cTTS and wTTS is of interest
because of the possibly different emission mechanisms: Next to the canonical interpretation
of magnetic structures with footpoints anchored on the star, cTTS may produce X-ray emission
related to the accretion process and/or in magnetic structures connecting the star with
the circumstellar disk. In wTTS the absence of a disk and accreting material imply that 
the most plausible explanation of the short-term X-ray variability is magnetic flaring.  

Observationally it is not straightforward to distinguish between variability originating from these
different causes, unless there is sufficient signal for detailed modelling of the spectral
evolution that gives access to the strength of the magnetic field and the length scale of 
magnetic structures \citep[see e.g.][]{Favata05.1}. A few of the brightest XEST sources are
examined in this respect by \citet{Franciosini06.1}. Here we compare the variability of
cTTS and wTTS using a statistical approach. 

In Sect.~\ref{sect:results_var} we have established that a similar fraction of cTTS and wTTS are
variable in the broad band. In the soft band, variability is more frequent
on wTTS than on cTTS: $18 \pm 6$\,\% variables among cTTS vs. $37 \pm 9$\,\% variable wTTS. 
A possible cause for this difference is the on average larger extinction of the former ones
that removes soft photons from the spectrum. About $30$\,\% of all cTTS have $A_{\rm V}<1$\,mag,
compared to more than two-thirds of the wTTS sample. 
This trend is even more pronounced among the stars that are constant in the soft band,
and supports the suspicion that soft absorption puts stronger limits on the detection of variability 
in cTTS than in wTTS. 
On the other hand, if a significant amount of X-rays from cTTS were
produced from accretion, then at least those cTTS that are not absorbed should show more -- not less -- 
soft variability than wTTS,
because the temperatures in a T Tauri accretion shock (of a few MK at most) 
can produce only soft photons ($<1$\,keV). This is not observed. 
An alternative explanation would be that accretion is steady
enough not to produce X-ray variability, or that variations in the density 
of the accreting material are smoothed out during the X-ray generation 
process \citep{Lamzin99.1}. 

\begin{figure}
\begin{center}
\resizebox{8cm}{!}{\includegraphics{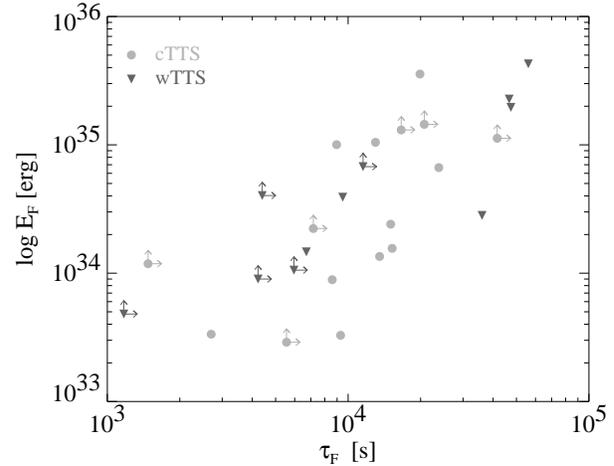}}
\caption{Flare energy and flare duration for cTTS and wTTS. Lower limits for flares that are unconstrained
because partially outside the observing window are indicated with arrows.}
\label{fig:flare_tau_energ}
\end{center}
\end{figure}

The XEST observations show further that flares (characterized by large amplitude and fast rise) 
are identified marginally more frequently on cTTS than on wTTS 
($31 \pm 7$\,\% vs $22 \pm 7$\,\%). 
A similar conclusion, but likewise not on a strong statistical basis, 
was obtained by \citet{Stelzer00.1} based on {\em ROSAT} observations. 
In Fig.~\ref{fig:flare_tau_energ} the flare durations and energies of cTTS and wTTS measured
during the XEST are shown. There is not a clear distinction between the two groups, pointing at a
similar -- most likely coronal -- origin for the X-rays from both classes of stars.  
%
%

\section{Conclusions}\label{sect:conclusions}

We have presented a systematic analysis of X-ray variability on pre-MS stars in the
TMC based on the XEST. This project includes a significant fraction of the known 
TMC members, and presents the deepest X-ray survey so far in 
the Taurus star-forming region. 
We have used an automated procedure for detecting variability and recognizing flares. 
Flares are defined with criteria that take into account the amplitude
and the derivative of the segmented lightcurves.
Variability is found in roughly half of the sample in the broad band from $0.3-7.8$\,keV. 
Our analysis in different energy bands has shown that it is important to include hard 
energies: Variability is detected more frequently in the hard band 
($50$\,\% of sources are variable at $>1$\,keV)
than in the soft band ($26$\,\% of sources are variable at $<1$\,keV) where, in some sources, 
strong extinction lowers the signal to the point that it becomes difficult to reveal variations.  
Next to this bias there is probably a physical reason behind the fact that variability 
shows up stronger at harder energies, related to the notion that variability goes 
along with additional heating. 

The detection of variability is biased because the Poisson statistics impose different
sensitivity thresholds for sources with different count rate, and the sample has a 
large range of count rates (four orders of magnitude). 
We have estimated the influence of this bias on the variability statistics, 
and evaluated the variability statistics for different classes of pre-main 
sequence stars (protostars, cTTS, wTTS, brown dwarfs), and for different spectral type ranges
that roughly correspond to mass for cool pre-MS stars. 

Variability is found slightly more
frequently on cTTS than on wTTS. The difference is only marginal, but when combined
with similar indications from various star-forming regions discussed in the 
literature \citep[e.g. ][]{Flaccomio06.1}, this finding may deserve further attention.
On the other hand, the flare energies and durations of cTTS and wTTS are similar,
and there is no evidence of extra soft variability from accretion in cTTS. All in all, the 
XEST data does not yield support for models that predict X-ray emission triggered
by reconnection related to star-disk magnetospheric interaction and accretion 
\citep[e.g. ][]{Shu94.1, Goodson99.1},
although such scenarios may apply in individual cases; see also discussion by 
\citet{Franciosini06.1}.  

There is no dependence of the fraction of
variable sources on spectral type (comparing K and M stars). 
\citet{Calvet04.1} summarized a correlation between mass accretion rate,
$\dot{M}$, and stellar mass, $M$, that is found to be $\dot{M} \propto M^{1.95}$.
If X-ray production was related to the accretion activity (e.g., through
accretion shock heating or through influence on the coronal magnetic structure),
then stronger accretion would be expected to produce stronger (absolute)
fluctuations, and therefore X-ray variability should be different not only between
accreting and non-accreting stars, but also between lower- and
higher-mass cTTS in the sample. Since M stars and K stars show no distinction
with regard to absolute variability, we find no evidence that the
accretion process significantly influences X-ray variability.
\citet{Preibisch05.1} arrived at similar conclusions for the ONC sample 
based on a comparison of the thermal properties of cTTS and wTTS.

We have evaluated the flare rates for the events on TMC members
identified in this work, and compared them to equivalent data for the ONC sample obtained by {\em Chandra}
within the COUP. In both cases the sample represents pre-MS stars, but the ONC study
concerns a narrow mass range ($0.9\,M_\odot < M < 1.2\,M_\odot$), 
while in this work we have included all TMC members detected during the XEST.  
The frequency of large flares ($E_{\rm F} > 10^{35}$\,erg) in the TMC 
is $1$ event per star in $770$\,ksec, 
roughly comparable to the estimate for the young solar analogs in the ONC. 

We have found twofold observational support for coronal heating by flares
in our stellar sample.  First, we have noted that the flare 
amplitude is correlated with the characteristic emission level of the
flaring star. In particular, we found an upper envelope to
the flare amplitudes as a function of $R_{\rm ch}$. While the
envelope is unlikely to be strict because longer observations
could detect some of the very rare but very large flares on any star,
the trend can naturally be explained if flares are related to the
coronal heating process \citep{Audard00.1}: Either, both the flare rate and
the overall heating are the result of a similar process, probably magnetic reconnection,
and therefore both the flare rate for a given flare amplitude and the total X-ray
luminosity are expected to scale with the coronal volume available for these processes.
Alternatively, the characteristic emission could be the direct result of the heating
process of a large number of unresolved flares occurring continuously on the star.
Both assumptions relate the physical processes relevant to flares to
the overall stellar radiative X-ray losses. In both cases, one expects
(for a given observing time) 
that the maximum of the absolute amplitude 
be proportional to $R_{\rm ch}$, and this is indeed what we find (Fig.~\ref{fig:rate_minmax}\,left).

Secondly, we have analyzed the cumulative number distribution of flare energies, making use of
our new results for the TMC and of the flare energies tabulated by \citet{Wolk05.1} for the ONC. 
Our ML estimates show that the distributions of both samples can be represented by a power-law, 
with an index of $\alpha = 2.4 \pm 0.5$ for the differential distribution of the TMC, and
$\alpha = 1.9 \pm 0.2$ for the ONC. 
\citet{Wolk05.1} had provided a value of $\alpha \sim 1.7$ for the same ONC data, using a linear
least-squares fit to the double logarithmic representation of the cumulative number distribution.
However, that approach does not take into account that the individual bins are not independent.
Therefore, our method is more rigorous. 
The power-law slope for the number distribution of flare energies in the TMC and in the ONC, 
although not well constrained, is in agreement with the range found 
in previous studies of solar and stellar flares (see references in 
Sect.~\ref{sect:flare_energy}),  
indicating that the energy release in cool stars is a fairly universal process. 
The observed values for $\alpha$ are near and probably larger than 
the critical limit ($\alpha_{\rm crit} = 2$) . 
Therefore, although the uncertainties preclude definite conclusions, micro-flare heating may well 
play an important role in heating these pre-MS coronae.

In summary, the entire sample supports
an important role of flare-like events in the generation of
the X-ray light curves, in a similar manner as observed before in
non-accreting magnetically active stars. In particular, 
we conclude that cTTS and wTTS 
reveal similar variability characteristics. These results point
at magnetic energy release playing the dominant role in the generation
of X-rays, while there is little evidence for an influence of accretion
disks (via long-scale magnetic fields) or accretion itself (via
accretion-shock heating). While neither of the latter two can be
excluded to contribute to the observed X-rays, we claim
that neither is responsible for strong X-ray losses, but that compact 
magnetic field annihilation dominates the X-ray production.  

\begin{acknowledgements}
We warmly acknowledge financial support by the International Space Science
Institute (ISSI) in Bern to the {\it XMM-Newton} XEST team. 
BS, EF, GM and IP acknowledge financial contribution from contract ASI-INAF I/023/05/0. 
X-ray astronomy
research at PSI has been supported by the Swiss National Science Foundation
(grants 20-66875.01 and 20-109255/1). MA acknowledges support by NASA grants
NNG05GF92G. BS thanks J.Bouvier for stimulating comments on the manuscript. 
This research is based on observations obtained with
{\it XMM-Newton}, an ESA science mission with instruments and contributions
directly funded by ESA member states and the USA (NASA). 
\end{acknowledgements}

\end{document}